\documentclass[graybox]{svmult}

\usepackage{mathptmx}       
\usepackage{helvet}         
\usepackage{courier}        
\usepackage{type1cm}        
\usepackage{makeidx}         
\usepackage{graphicx}        
\usepackage{multicol}        
\usepackage{lscape}
\usepackage[bottom]{footmisc}

\makeindex             

\usepackage[english,activeacute]{babel}
\usepackage{amsfonts,amssymb,amsmath}
\usepackage[a4paper, total={6in, 8in}]{geometry}
\usepackage{graphicx,epsfig}
\usepackage{color}
\usepackage{url}
\usepackage{forest}
\usepackage{multicol}
\usepackage{multirow}
\usepackage{caption}
\usepackage{subcaption}
\usepackage{multirow}
\usepackage{hyperref}
\hypersetup{colorlinks=false,pdfborder={0 0 0},}
\hypersetup{colorlinks,citecolor=black,filecolor=black,linkcolor=black, urlcolor=black}
\newcommand\BibTeX{{\rmfamily B\kern-.05em \textsc{i\kern-.025em b}\kern-.08em
T\kern-.1667em\lower.7ex\hbox{E}\kern-.125emX}}

         \def\UnderWiggleTemp{\the\catcode`\@}
         \catcode`\@=11
         \ifx\UnderWiggle@Loaded\relax
           \catcode`\@=\UnderWiggleTemp
           \endinput \else \let\UnderWiggle@Loaded=\relax \fi
         \newbox\U@BoxA
         \newbox\U@BoxB
         \newdimen\U@DimenA
         \def\U@DoUnderWiggle{
           \offinterlineskip
           \vtop{
             \hbox{\vbox{\copy0}}
             \vskip 1.2pt  
             \vbox to 0.4pt{
               \hbox to\wd0{\hss\char'176\hss}
               \vskip0pt minus 1fil
             }
             \vskip 0.4pt  
           }
         }
         \def\UnderWiggle#1{{%
           \ifmmode
             \mathchoice
               {\setbox0=\hbox{$\displaystyle #1$}\U@DoUnderWiggle}
               {\setbox0=\hbox{$\textstyle #1$}\U@DoUnderWiggle}
               {\setbox0=\hbox{$\scriptstyle #1$}\U@DoUnderWiggle}
               {\setbox0=\hbox{$\scriptscriptstyle #1$}\U@DoUnderWiggle}
           \else
             \setbox0=\hbox{#1}\U@DoUnderWiggle
           \fi
         }}
         \catcode`\@=\UnderWiggleTemp
         \newcommand{\uw}{\UnderWiggle}

\begin{document}

\title*{Utilizing Wavelet Transform in the Analysis of Scaling Dynamics for Milk Quality Evaluation}
\author{Devon Maywald and Dixon Vimalajeewa}
\institute{Devon Maywald  \at Department of Statistics, Texas A\&M University, College Station, Texas, USA \email{dmaywald@tamu.edu}
\and Dixon Vimalajeewa \at Department of Statistics, University of Nebraska Lincoln, Lincoln, Nebraska \email{hvimalajeewa2@unl.edu}}
%
%
\maketitle


\abstract{Food safety and quality are paramount concerns worldwide, especially concerning nutritional quality and its impact on human health. Ensuring the accuracy and efficiency of milk quality assessment is vital for maintaining the quality of dairy farm produce. Milk spectral data, Mid-infrared spectra (MIRS) of milk samples, are frequently employed for milk quality evaluations, encompassing various milk quality parameters. However, conventional milk quality analyses have overlooked the scaling nature, known as stochastic similarity in different scales, inherent in milk spectral data. Wavelet transforms are among the tools used in these analyses, although they are primarily used as data pre-processing techniques without fully realizing their potential in extracting valuable insights. The primary purpose of this study is to demonstrate the importance of accounting for scaling properties in assessing milk quality. A set of 12 descriptors is computed to characterize scaling properties in milk spectral data within the wavelet domain. These descriptors are then assessed for their effectiveness in milk quality assessments utilizing 18 different milk quality parameters. They notably demonstrated comparable performance to existing methods while utilizing fewer features when applied to an MIRS dataset. This innovative approach holds substantial promise for advancing the field of milk quality assessment, offering a means to achieve more accurate and efficient evaluations while shedding light on previously unexplored aspects of milk spectral data.}

\section{Introduction} \label{sec:1}
According to the United Nations Food and Agriculture Organization, there will be substantial challenges in meeting an increasing food demand due to a projected global population of 9.2 billion by 2050 \cite{Alexandratos2012}. This pressing need motivated the development of the concept of Smart Farming. As a paradigm of agriculture, Smart Farming leverages technology, data, and automation to improve efficiency, productivity, and sustainability. Among the focus points of smart dairy farming, analysis of milk quality stands out as a crucial component of meeting future food demands. 

 The assessment of milk quality generally entails the analysis of its composition, such as fatty acids, proteins, coagulation characteristics, acidity characteristics, minerals, and geographical origin \cite{Su2019}. This also helps optimize individual animal traits relating to health, nutrition, well-being, and environment. Mid-infrared (MIR) spectroscopy has emerged as the most economically viable method in these analyses. It uses infrared radiation to measure the chemical composition of liquid foods such as milk. Additionally, near-infrared spectroscopy (NIRS), optical analysis, milk conductivity analysis, and microstrip sensors are also used \cite{Kunes2021, Amar2020}. Overall, a comprehensive assessment of milk composition is essential not only for determining its quality but also for optimizing a variety of aspects of dairy farming, including product safety, nutrition, animal welfare, and sustainable farming methods.

Data analysis tools are crucial for processing and extracting insights about milk composition from large milk spectral datasets. Deep learning (DL) is one of the popular tools in the literature \cite{Helan2023, Vidhya2023, Dixon2022a}. Most of these existing tools primarily rely on wavenumbers corresponding to peak intensity values in MIRS and principal component analysis of MIRS \cite{Frizzarin2020, Dixon2018}. Given their model complexity, intricate nature, and high dimensionality of milk spectral data, wavelet-based tools have been proven suitable for building effective milk spectral data analysis tools. \cite{Dixon2017, Dixon2018}. However, wavelet-driven methods are primarily applied to feature extraction, data compression, or denoising in this context \cite{Tsenkova2022, Dixon2017, Dixon2018}, overlooking other valuable properties inherent in milk spectral data that wavelets can extract. 

Scaling nature is one such property. It signifies the stochastic similarity observed within signals at varying resolutions (or scales). Specifically, scaling behavior characterizes intermittency (or randomness) in signal dynamics. It has been extensively demonstrated in other application domains (e.g., medicine and engineering) that the analysis of scaling nature can uncover valuable information that standard statistical measures, such as mean and variance, fail to reveal \cite{DixonOvarian2023, Dixon2022a}. For instance, lower intermittency in a signal is attributed to its higher correlation and long-memory \cite{Dixon2023c}.
However, explorations of scaling nature in milk spectral data have been limited. Therefore, the key motivation of this study is to demonstrate the importance of accounting scaling nature in milk spectral data analysis. Thus, the main contribution of this study is providing a novel perspective, particularly for milk quality analysis, proposing a method solely based on scaling properties analyzed in the wavelet domain. Therefore, this method differs from existing methods, which primarily rely on wavenumbers corresponding to peak intensity values in MIRS or principal component analysis of MIRS \cite{Frizzarin2020, Dixon2018}.

This study focuses on introducing a set of descriptors by exploring scaling behaviors in milk spectral data characterized in the wavelet-domain. Monofractality and multifractality, two aspects of scaling behaviors, are considered when computing these descriptors. Monofractality can be characterized using a single irregularity index, a constant between 0 and 1 that quantifies signal intermittency (more details in Section \ref{sec:Methods}). Multifractality provides insight into signal inhomogeneity by encapsulating the diversity of irregularity indices. As part of analyzing monofractal properties, the wavelet spectrum is used to assess signal scaling behavior at different resolutions. This approach characterizes the level-wise decay in scale-specific average "energies" of the wavelet coefficients obtained from the wavelet decomposition. The term "energy" used here quantifies the magnitude of squared wavelet coefficient. The rate of this energy decay along increasing scales (or resolutions) quantifies the degree of regularity in milk spectral data. A wavelet-based multifractal spectrum is computed, and its properties are explored to assess multifractality. Across different signal scales, the multifractal spectrum describes the distribution of irregularity indexes. Twelve descriptors are established using the mono- and multifractal spectra of milk spectra. Finally, their performance is examined to determine variability in three milk quality parameter groups, including 18 milk quality traits in total, using MIRS data consisting of 712 MIR spectra. 

The remainder of the paper is organized as follows. The techniques used in this study, fundamentals of wavelet transform, and assessment of monofractal and multifractal properties by using wavelet transforms are presented in Section \ref{sec:Methods}. Sections \ref{sec:MotivatingStudy} and \ref{sec:Results} provide data analysis procedures and results, respectively. Section \ref{sec:Discussion}, which discusses the results, is followed by some concluding remarks in Section \ref{sec:Conclusions}.

\section{Methods}\label{sec:Methods}
This section describes the procedures used to compute scaling descriptors by using wavelet transforms to assess the scaling nature in milk spectral data. We briefly introduce wavelet transforms and describe how scale properties in milk spectral data can be assessed using monofractals and multifractals.

\subsection{Wavelet Transform}\label{sec:wt}
As a signal processing tool, wavelet transforms (WTs) decompose a signal into contributions that are localized both in time and frequency. Doing so makes it possible to analyze the signal simultaneously at various resolutions or scales by presenting it in the wavelet domain in a hierarchical manner. This allows WTs to reveal signal properties that are otherwise difficult to determine. 

The Discrete Wavelet Transform (DWT) is a commonly used type of WT that has become a popular tool for analyzing complicated signals in application domains where discrete data is analyzed. DWTs are linear transforms that have the following mathematical representation.

Suppose a data vector $Y = (y(t_1), y(t_2), \cdots, y(t_N))'$  of size $N \times 1$ is measured at $N$ equally spaced points $t_i$ for $i = 1, 2, \cdots, N$. The DWT of  $Y$  is given by
\begin{equation}\label{eqn-12}
  d = WY,
\end{equation}
where $d$ is also a vector of size $N \times 1$, and $W$ is a wavelet-specific orthogonal matrix of size $N \times N$. The elements in $W$ are determined by wavelet filter corresponding to the selected wavelet basis, such as the Haar, Daubechies, or Symmlet families~\cite{vidakovic1999}.

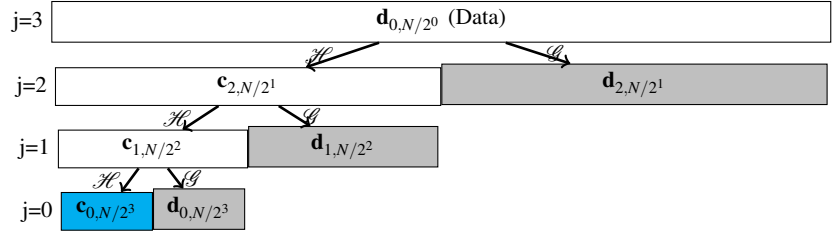
\begin{figure}[!t]
\center
    \begin{forest}
    for tree = {grow=south,
      l sep=0.0cm,
      s sep=0.0cm,
      minimum height=0.5cm,
      minimum width=1cm,
      edge={->,line width=1pt},
      text centered,
      }
    [$\mathbf{d}_{0,N/2^0}$ (Data), draw, text width=10.05cm, label={left:{j=3}}
      [$\mathbf{c}_{2,N/2^1}$, draw, text width=4.87cm, edge label={node[midway,left]{$\mathcal{H}$}}, label={left:{j=2}}
       [$\mathbf{c}_{1,N/2^2}$, draw, text width=2.29cm, edge label={node[midway,left]{$\mathcal{H}$}}, label={left:{j=1}}
         [$\mathbf{c}_{0,N/2^3}$, draw, text width=1cm, fill=cyan, edge label={node[midway,left]{$\mathcal{H}$}}, label={left:{j=0}}]
         [$\mathbf{d}_{0,N/2^3}$, draw, text width=1cm, fill=lightgray, edge label={node[midway,right]{$\mathcal{G}$}}]
       ]
       [
       $\mathbf{d}_{1,N/2^2}$, draw, text width=2.29cm, fill=lightgray, edge label={node[midway,right]{$\mathcal{G}$}}
       ]
      ]
      [$\mathbf{d}_{2,N/2^1}$, draw, text width=4.87cm, fill=lightgray, edge label={node[midway,right]{$\mathcal{G}$}}
        ]
      ]
    ]
  \end{forest}
  \caption{A schematic representation of the discrete wavelet transforms (DWTs). DWT is performed three times on a signal of length $N (= 2^J)$. At each decomposition level, $j$, the coefficients $c$  denotes smoothing coefficients obtained via low-pass filter $H$ while $\uw{d}$ denotes the detail wavelet coefficients obtained via high-pass filter $G$.}
\label{fig:DWT_Method}
\end{figure}

Utilizing the matrix-based approach for the DWT, as shown in (\ref{eqn-12}), becomes computationally intensive when dealing with a large value of $N$. To address this, a computationally efficient technique, pioneered by Mallat, has been employed for performing DWT on datasets with sample sizes that are powers of 2 (e.g., $N = 2^J$, where $J \in \mathbb{Z}^{+}$) \cite{Mallat1989}. This algorithm follows a hierarchical structure. It involves a series of convolutions carried out using a wavelet-specific low-pass filter and its corresponding high-pass filter. Subsequently, a process of decimation (selecting values at even positions within a sequence) is applied. As a result, a multiresolution representation of the signal is obtained, consisting of a smoothed approximation ($\uw{c}$) and an array of detail coefficients $d_{jk}$ at varying resolution levels (indexed by $j$) and positions ($k$) within each resolution level. Thus, for a vector $Y$ of size $N=2^J$, the vector $d$ in (\ref{eqn-12}), has the following structure

\begin{equation}
\label{eq:wd}
     d = (\uw{c}_{J_0}, \uw{d}_{J_0}, \dots, \uw{d}_{J-2}, \uw{d}_{J-1}),
\end{equation}
where $\uw{c}_{J_0}$ is a vector of coefficients corresponding to a smooth trend in the signal, $\uw{d}_{j}$ are detail coefficients at different resolutions $j$,     and $J_0$ is the coarsest resolution level in the wavelet decomposition, such that $1 \leq J_0 \leq J-1$. Vector $d$ in (\ref{eq:wd}) has $J_0$ levels of detail coefficient vectors,  $\uw{d}_{J_0}, \dots, \uw{d}_{J-2}, \uw{d}_{J-1}$, which are used in the definition of wavelet spectra. For instance, Figure \ref{fig:DWT_Method} illustrates the DWT when $J_0 = 3$. This algorithm is now available in many standard wavelet packages (e.g., \href{https://statweb.stanford.edu/~wavelab/}{\emph{WAVELAB}}  module for MATLAB ). Interested readers can find more details about WTs and different applications of WTs in \cite{vidakovic1999}.

\begin{figure*}[!t]
\centering
    \vspace{-1.25em}
    \includegraphics[width= .7\linewidth]{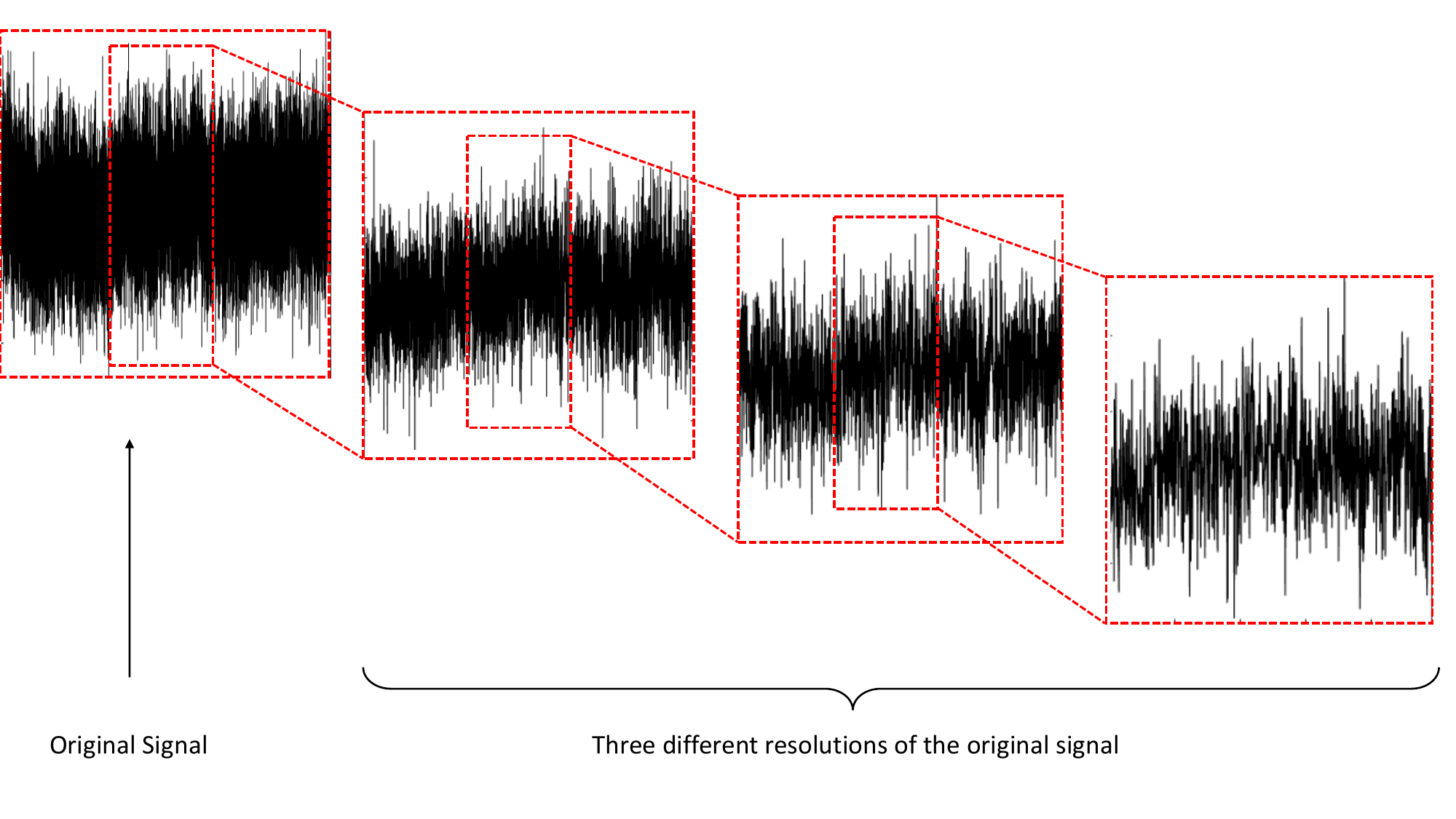}
    \caption{ A sample high-frequency signal at three different resolutions. The signal exhibits similar properties and behaviors (e.g. mean, variance)  when exploring the signal at different resolutions. The tendency to exhibit such similar properties at different resolutions is identified as the self-similar nature of the signal.}
    \label{fig-00a}
\end{figure*}

\subsection{Extraction of Scaling Descriptors}\label{sec:fract}
Scaling behavior is generally characterized by fractality, which can either manifest as a monofractal or a multifractal pattern. Rather than discussing theoretical information, we describe the process of computing scaling descriptors. Interested readers may obtain theoretical details in \cite{Goncalves1998, vidakovic1999, Dixon2023}.

\subsubsection{Assessing Monofractal Properties}\label{sec:self-similar}
Monofractality pertains to the characteristic of a signal where its scaling attributes maintain consistency across varying scales. This tendency is commonly noticed in signals possessing uncomplicated and orderly structures. In such instances, the scaling properties persist across all scales uniformly, and the system displays a sole fractal dimension. 

\begin{figure}[!t]
\centering
\begin{subfigure}{.55\linewidth}
  \centering
  \includegraphics[width=.8\textwidth]{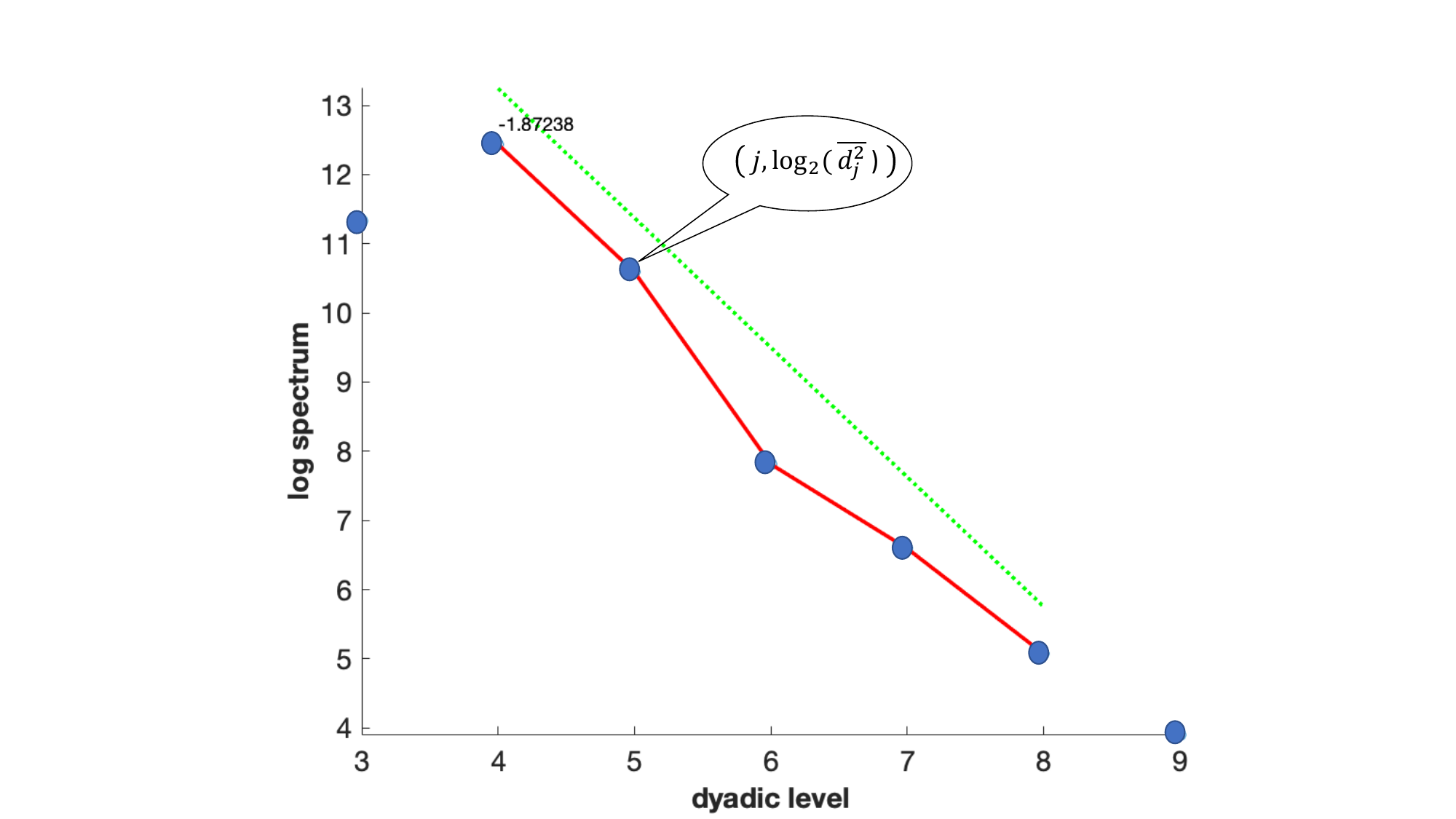}
  \caption{}
  \label{Mofrac}
\end{subfigure}
\begin{subfigure}{.425\linewidth}
  \centering
  \includegraphics[width= .8\textwidth]{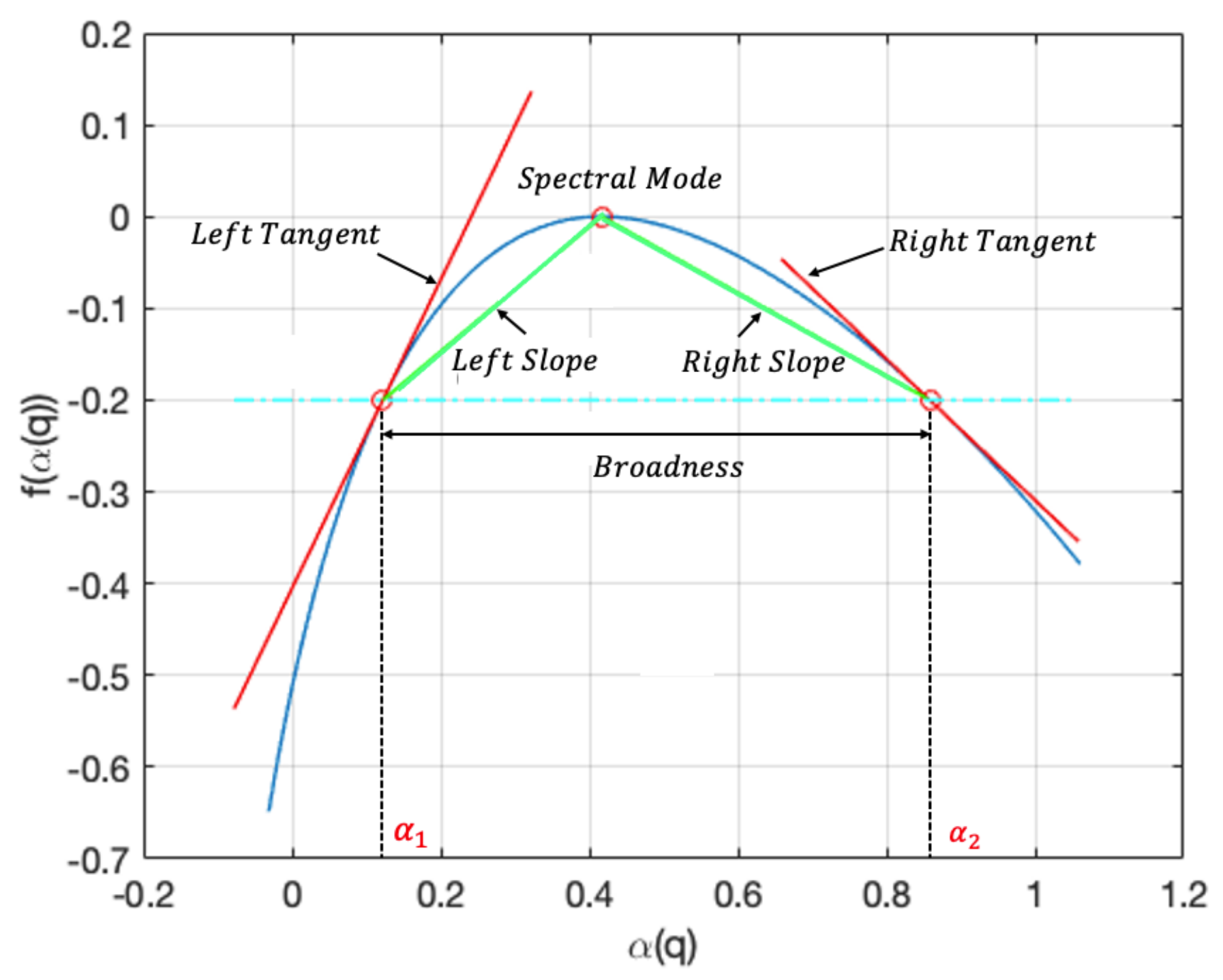}
    \vspace{-0.75cm}
  \caption{}
  \label{Mfrac}
\end{subfigure}
\caption{Assessment of scaling properties: (a) A sample monofractal wavelet spectra. Slope of the wavelet spectra is estimated by fitting a straight line (green dashed) on the log energy of the wavelet coefficients (black) within the scale index ranging from 1 to 10 (red line). The coordinate of the point at the level $j$ is $\log_2 \left( \bar{\uw{d}_j^2} \right)$, where $\uw{d}_j$ is the wavelet coefficients at the scale index j and (b) Multifractal spectrum and its geometric descriptors; $x$-axis represents the irregularity index ($H$), $\alpha(q)$ and y-axis represents values proportional to the relative frequency of $H$, $f(\alpha(q))$.}
\label{fig-111}
\end{figure}

In the wavelet-domain, the wavelet spectrum is commonly employed to characterize the scaling tendencies of the signal. This spectrum is generated by taking the logarithmic average of the squared detail wavelet coefficients, also known as log energies, across various scales. Signals exhibiting scaling behavior (or self-similarity) manifest a distinct pattern in their wavelet spectra: the log energies exhibit a linear decline as resolution decreases or scale increases (refer to Figure \ref{Mofrac}). The rate of this energy reduction, determined by regressing the log energies against the scale indices, quantifies the level of regularity. This is typically denoted by the Hurst exponent, $H$ (aka irregularity index), which satisfies the equation $H = -(slope + 1)/2$. Higher slopes ($> -2$) indicate a heightened degree of persistence, while smaller slopes ($< -2$) denote greater antipersistence and intermittency. The standard Brownian motion serves as a critical reference point with a log spectrum characterized by a slope of -2. For detailed information, we refer to \cite{vidakovic1999}.

\begin{table*}[!t]
\centering
\caption{ The set of scaling descriptors computed in the wavelet domain of the MIRS (see Figure \ref{fig-111} for graphical representation of these descriptors).}
\begin{tabular}{|c|l|l|}
\hline
{\bf } & {\bf Name } & {\bf Description } \\
\hline
\multicolumn{3}{|c|}{ {\bf Based on Monofractal Behavior} } \\
\hline
1 & Slope (S)  & The rate of energy decay in wavelet spectrum  \\
\hline
\multicolumn{3}{| c |}{ { \bf Based on Multifractal Behavior} } \\
\hline
\multirow{3}{*}{2} & \multirow{3}{*}{Left Slope (LS)} & Slope of the distribution produced by the\\
&& collection of regularity index $\alpha$ with smaller\\
&& values of the spectral mode\\
\hline
\multirow{3}{*}{3} &  \multirow{3}{*}{Right Slope (RS)}  & Slope of the distribution produced by the \\
&& collection of regularity index $\alpha$ with larger \\
&& values of the spectral mode\\
\hline
4 & Broadness (B) & Absolute difference between $\alpha_1$ and $\alpha_2$\\
\hline
\multirow{2}{*}{5} &  \multirow{2}{*}{Spectral Mode (SM)}  &   Most frequently occurring regularity index $\alpha$ \\
&&found within the signal\\
\hline
6 &  Left Tangent Point (LTP)  & The point $ f(\alpha_1)$ on the spectrum\\
\hline
7 &  Right Tangent Point (RTP)  & The point $ f(\alpha_2)$ on the spectrum\\
\hline
8 &  Left Tangent (LT)  &  Tangent at the point $(\alpha_1, f(\alpha_1))$ on the spectrum\\
\hline
9 &  Right Tangent (RT)  &  Tangent at the point $(\alpha_2, f(\alpha_2))$ on the spectrum.\\
\hline
\multirow{2}{*}{10} & \multirow{2}{*}{Max Curvature (MC)} & The evaluation of the curvature formula (\ref{eq:curvature}) \\ && applied to $(\alpha, f(\alpha))$ at the peak of $(\alpha, f(\alpha))$ \\
\hline
\multirow{2}{*}{11} & \multirow{2}{*}{Curvature (K)} & Average of lower order approximations of $|f''(\alpha)|$ \\ && where $f(\alpha)$ is maximal \\
\hline
\multirow{2}{*}{12} & \multirow{2}{*}{Curvature Central (KC)} & Higher order approximation of $|f''(\alpha)|$ \\ && where $f(\alpha)$ is maximal\\
\hline
\end{tabular}
\label{tab-0}
\end{table*}

\subsubsection{Assessing Multifractal Properties}\label{sec:Multifractal}
In contrast to monofractal signals, a signal exhibits multifractality when its structure demonstrates varying scaling properties across different scales. The multifractal spectrum serves as a potent tool for investigating these multifractal characteristics. It portrays the relative richness of diverse irregularity indices within the signal. To elaborate, constructing the multifractal spectrum involves the computation of local singularity strength or the Hurst exponent at individual points within the signal. Subsequently, the distribution of these computed values across varying scales is measured. For a more in-depth understanding of the methodology to calculate the multifractal spectrum in the wavelet domain, detailed technical information can be found in the work by Goncalves et al. \cite{Goncalves1998}

A number of meaningful descriptors can be used to describe the multifractal spectrum of a signal instead of considering it as a function. The multifractal spectrum, despite its functional nature, can generally be encapsulated by three key summaries: \emph{spectral mode (SM)}, \emph{broadness (B)}, and \emph{left (or right) tangent (LT/RT)} (depicted in Figure \ref{Mfrac}). The \emph{spectral mode} represents the most frequently occurring scaling index within the spectrum. In the context of monofractal signals, this spectral mode aligns with the Hurst exponent ($H$). The \emph{broadness} (or bandwidth) constitutes a more intricate descriptor of the multifractal spectrum. It is calculated as the absolute difference between $\alpha_1$ and $\alpha_2$, where $f(\alpha_1) = f(\alpha_2) = a$ for a chosen value of $a$ (typically around $-0.2$ in practical scenarios). A larger broadness value signifies the existence of multiple degrees of scaling parameters within the signal, indicating a departure from monofractality. The \emph{left tangent} corresponds to the slope of the tangent line at the point $(\alpha_1, f(\alpha_1))$ on the spectrum. Similarly, the \emph{right tangent} reflects the tangent slope on the opposite side. These tangent descriptors also serve as indicators of deviation from monofractality. Specifically, spectra with a lower slope of the left tangent indicate the presence of multifractality, as a pure monofractal process theoretically possesses an infinite LT \cite{Goncalves1998}. 

Three additional descriptors, max curvature, curvature, and average curvature, are developed by utilizing the curvature of the multifractal spectrum. According to \cite{Molina2019}, at any point along the multifractal spectrum, its curvature is defined as 

\begin{equation}
\label{eq:curvature}
     \text{Curvature}[(\alpha, f(\alpha))] = \frac{|\frac{d^2f}{d\alpha^2}|}{[1 + (\frac{df}{d\alpha})^2]^{3/2}}.
\end{equation}

In this study, the scaling descriptor \emph{max curvature} is calculated by evaluating the curvature of the multifractal spectra at its peak. This formula is complicated to calculate, especially considering that $\alpha$ and $f$ are functions of the moment orders ($q$). Many numerical derivatives are needed to evaluate (\ref{eq:curvature}), as well as $\alpha$ and $f(\alpha)$ themselves. To supplement the scaling descriptor \emph{max curvature}, the \emph{curvature} and \emph{curvature central} scaling descriptors are calculated by simplifying the curvature formula to just $f(\alpha)$ itself following the study in \cite{Molina2019}. By noting that $f'(\alpha) = 0$ at the maximal value of $f(\alpha)$, evaluating the curvature of $f(\alpha)$ at the peak involves only calculating $|f''(\alpha)|$. The scaling descriptor \emph{average curvature} takes the average of different lower order approximations of $|f''(\alpha)|$ where $f(\alpha)$ is maximal. The scaling descriptor \emph{curvature central} differs by taking a higher order approximation of $|f''(\alpha)|$.

Overall, the monofractal and multifractal spectra provide a set of descriptors to characterize complex behaviors of signals. In addition to describing the self-similarity and irregularity of signals, these spectra also offer information about their inhomogeneity. Table \ref{tab-0} summarizes the descriptors derived from both the spectrum and their graphical representation are depicted in Figure \ref{fig-111}.

\section{Milk Spectral Data Analysis}\label{sec:MotivatingStudy}
Mid-infrared (MIR) spectroscopy can identify the composition in liquid foods through their ability to absorb frequencies characteristic of their structure. A unique vibrational energy is associated with each chemical bond in a molecule, indicating that each compound has a unique fingerprint to identify its structure \cite{Su2019}. Therefore, the data obtained by the MIRS can be considered the fingerprint of the chemical bonds present in the molecules of a milk sample. MATLAB was used to compute scaling descriptors, and classification models were implemented using Python.

\begin{figure*}[!t]
\centering
    \includegraphics[width= .7\linewidth]{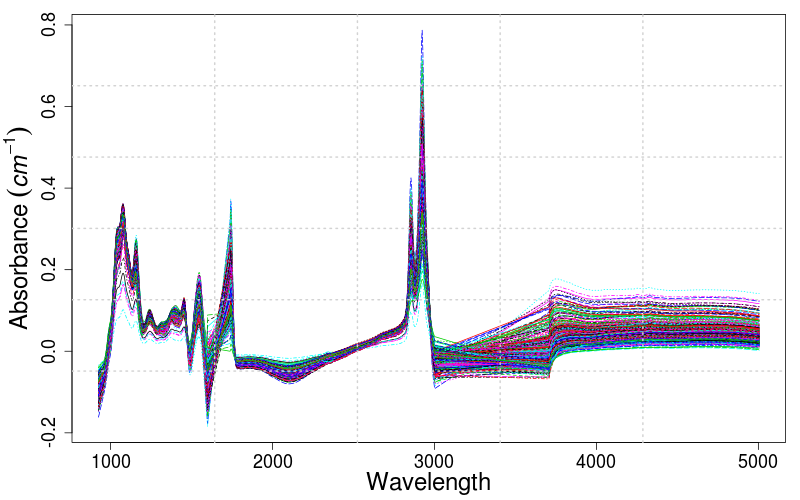}
    \caption{ The MIRS dataset. This data contains 712 milk spectra collected from 605 cows over a period of 35 days.}
    \label{MIRS_data}
\end{figure*}

\subsection{MIRS Dataset}\label{pre_process1}
The data utilized in this study was collected from the Teagasc research dairy farm in Moorepark, Ireland. The data collection spanned one year, commencing in August 2013 and concluding in August 2014, and involved 605 dairy cattle. Milk composition was determined using FT-MIR technology, specifically the FOSS MilkScan prediction equations. The dataset consisted of the spectra of 712 unique milk samples, covering the wavenumber range of $925-5005 cm^{-1}$ with a resolution of $3.853 cm^{-1}$. Each MIR milk spectrum is comprised of 1060 transmittance data points. To facilitate analysis, the spectral values, initially provided as transmittance, were converted to absorbance by computing the logarithm (base 10) of the reciprocal of the transmittance values (refer to Fig.~\ref{MIRS_data}). 

\subsection{Milk Quality Parameters}\label{pre_process}
We investigated three milk quality parameter (MQP) groups: technological, protein, and general traits. It is important to note that the selection of these specific milk traits aimed to enable a fair comparison between machine learning models based on the proposed scaling descriptors and existing models built using other descriptors, particularly principal component analysis (PCA) and direct wavelengths in the spectra. For instance, similar studies based on the same MIRS dataset have employed those PCA descriptors in \cite{  Dixon2017, Dixon2018} and wavelength descriptors in \cite{ Frizzarin2020}.

\begin{table*}[!t]
\centering
\caption{The set of milk quality parameters explored using the proposed scaling descriptors  }
\begin{tabular}{|c|l|c|c|}
\hline
{\bf } & {\bf Milk Quality Parameter} & {\bf Abbreviation} & {\bf Number of Samples} \\
\hline
\multicolumn{4}{|c|}{ {\bf Technological traits} } \\
\hline
1 & Rennet coagulation time  & RCT  & 484\\
\hline
2 & Curd-firming time at 30 minutes & k20 & 449\\
\hline 
3 & Curd firmness at 30 minutes  & a30 & 421 \\
\hline 
4 & Curd firmness at 60 minutes & a60 & 480 \\
\hline 
5 & Casein micelle size & CMS  & 585 \\
\hline 
6 & Native pH & pH & 596 \\
\hline 
7 & Heat Stability & HS & 422 \\
\hline 
\multicolumn{4}{| c |}{ { \bf Milk Protein Traits} } \\
\hline
1 & $\alpha_{S1}$-casein & $\alpha_{S1}$-CN & 471 \\
\hline
2 & $\alpha_{S2}$-casein & $\alpha_{S2}$-CN & 473 \\
\hline 
3 & $\beta$-casein & $\beta$-CN & 471 \\
\hline 
4 & $\kappa$-casein & $\kappa$-CN & 471 \\
\hline 
5 & $\alpha$ lactoglobulin  & $\alpha$-L & 472 \\
\hline 
6 & $\beta$ lactoglobulin  A & $\beta$-LGA & 472 \\
\hline
7 & $\beta$ lactoglobulin B  & $\beta$-LGB & 434 \\
\hline
\multicolumn{4}{| c |}{ { \bf General Milk Quality Traits} } \\
\hline
1 & Total lactose content &  TLC & 600 \\
\hline
2 & Total urea content &  TUC & 600 \\
\hline
3 & Total fat content &  TFC & 579 \\
\hline
4 & Total protein content &  TPC & 601 \\
\hline
\end{tabular}
\label{tab-01}
\end{table*}

Missing data within the milk traits have been recorded with zeros during the data collection process. Such observations were removed from the data in the present study. Also, since multiple samples were recorded for some animals (i.e., 712 spectra from  605 cows), to have one sample per animal, where multiple observations existed, the average MIRS, average scaling descriptor, and average milk traits were taken. The MIRS data is then standardized on each transmittance value. 

Outliers for milk traits were considered to be those greater than three standard deviations away from the mean. Such observations were also removed from the dataset. Table \ref{tab-01} provides an overview of these specific milk traits and the resulting spectra per MQP.

The data was collected as continuous measurements, so recorded \emph{RCT, k20, a30, a60, CMS, pH, HS, TLC, TUC, and TFC} were categorized into two categories based on the median, namely $Q_{Low}$ and $Q_{High}$. However, the protein traits $\alpha_{S1}$-CN, $\alpha_{S2}$-CN, $\beta$-CN, $\kappa$-CN, $\alpha$-LA, $\beta$-LGA, and $\beta$-LGB as well as TPC were separated into four categories based on quartiles, namely $Q_1, Q_2, Q_3,$ and $Q_4$. These classifications by median and quartiles were employed to replicate the process used in \cite{Frizzarin2020}.

\subsection{Computing Scaling Descriptors and Identifying Their Significance}
We performed DWTs on the first 1024 ($=2^{10}$) transmittance values in each MIR spectrum. This was followed by the computation of monofractal and multifractal spectra of each MIR spectrum for computing the twelve scaling descriptors listed in Table \ref{tab-0}. 

Before building classification models, we determined the most significant scaling descriptors for each MQP based on their statistical significance for separating categories within the MQPs. For this analysis, kernel density estimates (KDE) of the proposed scaling descriptors were computed with respect to categories within each MQP. The two sample-two-sided Kolmogorov–Smirnov test was then performed on the KDE of a given descriptor to determine if the distributions differ between MQP categories. Binary classification models implement scaling descriptors in order of the reported p-value. 

Whenever the milk quality parameter is categorized by quartile, resulting in four categories, each scaling descriptor has six different Kolmogorov–Smirnov tests to perform. The scaling descriptors were ordered by their average reported p-value across these six tests. 

\subsection{Classification Models and Analyses}\label{sec: classification models}
 The classification models explored in this study include Support Vector Machine (SVM), Linear Discriminant Analysis (LDA), Gradient Boosting Machine (GBM), K Nearest Neighbors (KNN), Logistic Regression (Logit), Gaussian Naive Bayes (GNB), Quadratic Discriminant Analysis (QDA), and Partial Least Squares Discriminant Analysis (PLSDA). 
 
 When fitting a classification model, scaling descriptors were implemented successively in order of statistical `significance', i.e., in order of their reported p-value from the Kolmogorov–Smirnov tests. A demonstration of this procedure can be seen in Figure \ref{Scaling Descriptor KDE Example}. The order of scaling descriptors implemented for each milk quality parameter is summarized in Table \ref{Best_descriptors} provided in the Appendix. 
 
 In model training and testing, we randomly assigned 80\% of the data samples to the training set and used the remaining samples for testing. A stratified split was done to preserve the balance of classes within the training and testing sets. Bayesian optimization was implemented with 10-fold cross-validation to determine optimal tuning parameters, as described in Table \ref{Classification_Method_Table}. Finally, classification performance was evaluated by computing classification accuracy. This procedure was repeated four times, and the average training and testing accuracy and standard deviation were calculated. Additionally, a comparison of classification performance was conducted using MIRS features computed using PCA. 
 
\section{Results}\label{sec:Results}

This section provides MIRS data analysis outcomes. First, we report variability in fractal (self-similar or scaling) properties as in their $mean \pm std$ for each milk quality parameter. This is followed by the classification performance of the scaling descriptors and PCA served from the MIRS dataset.  

In the Appendix, Tables \ref{Scaling_descriptors_summaryStats_1}, \ref{Scaling_descriptors_summaryStats_2}, and \ref{Scaling_descriptors_summaryStats_3}  offer a comprehensive overview of the variability in scaling properties. A detailed examination of the scaling property slope (S) for Milk Quality Parameters (MQPs) in two categories, $Q_{Low}$ and $Q_{High}$, as shown in Table \ref{Scaling_descriptors_summaryStats_1}, reveals that most MQPs have a steeper slope in the $Q_{Low}$ category compared to the $Q_{High}$ category. Interestingly, the opposite trend is observed in the group of milk quality traits summarized in Table \ref{Scaling_descriptors_summaryStats_3}. Furthermore, when we consider the broadness (B) of the same MQPs, the $Q_{High}$ category exhibits a higher B value compared to the $Q_{Low}$ category. This suggests that milk spectra with a higher proportion of technological traits and milk quality traits tend to be more irregular and inhomogeneous.

Upon analyzing the variability in scaling properties of MQPs related to milk protein across four quartiles, as presented in Table \ref{Scaling_descriptors_summaryStats_2}, we generally observe a higher slope and a lower broadness for the higher quartiles. The left slope of the MQPs displays a behavior similar to the broadness. This indicates that as the protein composition increases, the multifractality (or complexity and inhomogeneity) in milk spectral data also increases.

\begin{figure*}[!t]
\centering
    \includegraphics[width= 1\textwidth]{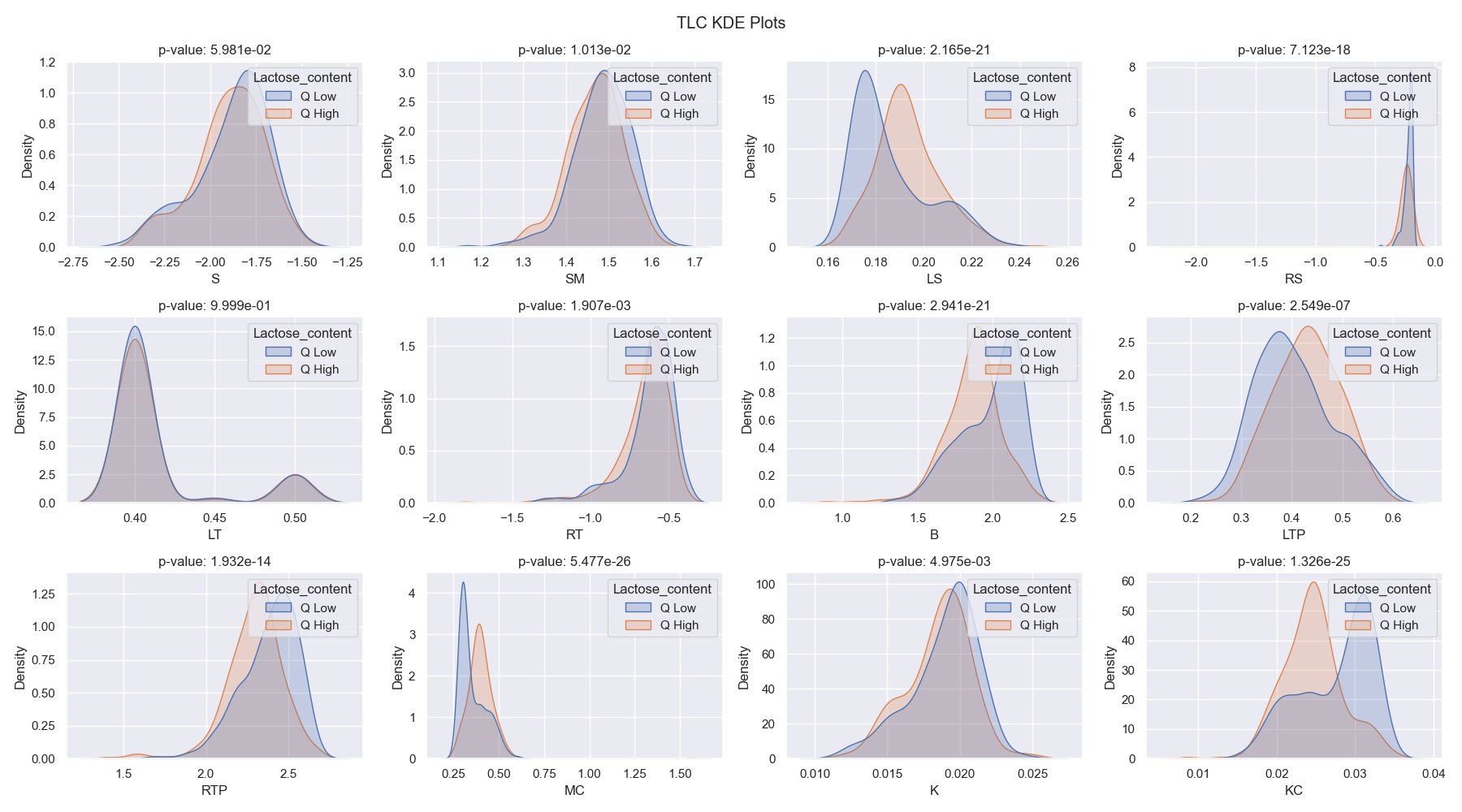}
    \caption{ Kernel density estimate of the scaling descriptors separated by classification of total lactose content (Q Low and Q High) and corresponding p-values from a two-sample two-sided Kolmogorov–Smirnov test performed on the separated KDE's. Models attempting to classify Total Lactose Content, TLC, based on scaling descriptors will additively implement scaling descriptors in the order lowest to highest p-value.}
    \label{Scaling Descriptor KDE Example}
\end{figure*}

\begin{table}[!t]
\centering
\caption{The scaling descriptors in order of reported p-value from the Kolmogorov–Smirnov test.}

\begin{tabular}{|c|c|l|}
\hline
{\bf MQP} & {\bf Best Model} & {\bf Scaling Descriptors} \\
\hline
\multicolumn{3}{|c|}{ {\bf Technological traits} } \\
\hline
RCT & Logit & KC, MC, LS, RTP, B, RS, SM, K, RT, S, LTP, LT \\
\hline
k20 & KNN & RTP, MC, KC, B, LS, RS, SM, K, RT, S, LTP, LT \\
\hline 
a30 & Logit & RTP, KC, MC, B, LS, RS, SM, K, RT, S, LTP, LT \\
\hline 
a60 & PLSDA & MC, KC, LS, B, RS, RTP, RT, LTP, K, SM, S, LT \\
\hline 
CMS & PLSDA & B, LTP, S, LS, MC, RT, KC, RS, K, RTP, SM, LT \\
\hline 
pH & GBM & MC, KC, SM, B, RTP, LS, RS, S, LT, K, LTP, RT \\
\hline 
HS & PLSDA & MC, KC, LS, B, RS, RTP, K, LTP, S, SM, RT, LT \\
\hline 
\multicolumn{3}{| c |}{ { \bf Milk Protein Traits} } \\
\hline
$\alpha_{S1}$-CN & LDA & MC, KC, B, LS, RTP, RS, SM, S, RT, LTP, K, LT \\
\hline
$\alpha_{S2}$-CN & LDA & B, K, MC, KC, LS, RTP, RS, LTP, SM, S, RT, LT \\
\hline 
$\beta$-CN & SVM & LS, B, RS, KC, MC, RTP, S, K, LTP, LT, SM, RT \\
\hline 
$\kappa$-CN & Logit & LS, RTP, KC, MC, B, RS, K, RT, SM, LTP, S, LT \\
\hline 
$\alpha$-L & SVM & MC, B, LS, RS, RTP, KC, SM, S, LTP, LT, K, RT \\
\hline 
$\beta$-LGA & SVM & LS, MC, RTP, B, KC, RS, LTP, K, SM, RT, S, LT \\
\hline
$\beta$-LGB & SVM & LS, MC, RTP, KC, B, LTP, RS, SM, S, K, RT, LT \\
\hline
\multicolumn{3}{| c |}{ { \bf Milk Quality Traits} } \\
\hline
TLC & Logit & MC, KC, LS, B, RS, RTP, LTP, RT, K, SM, S, LT \\
\hline
TUC  & PLSDA & MC, KC, B, RS, LS, RTP, SM, RT, K, LTP, S, LT \\
\hline
TFC & SVM & S, LTP, SM, K, RTP, LT, LS, RS, B, KC, MC, RT \\
\hline
TPC & SVM & MC, KC, B, RS, LS, RTP, SM, K, LTP, RT, S, LT \\
\hline
\end{tabular}
\label{Best_descriptors}
\end{table}

Generally, both monofractal and multifractal scaling properties exhibit varying degrees of change across different milk quality parameters. To illustrate this process, Figure \ref{Scaling Descriptor KDE Example} shows KDEs of the scaling descriptors along with the p-values of the two sample-two-sided Kolmogorov–Smirnov test. The particular MPQ in the illustration is Total Lactose Content, TLC, after being separated into low and high categories (based on median). Table \ref{Best_descriptors} summarizes the order of significance (discriminatory level) of scaling descriptors implemented for each Milk Quality Parameter.

\begin{figure}[!t]
\centering
\begin{subfigure}{.475\linewidth}
  \centering
  \includegraphics[width=1\textwidth]{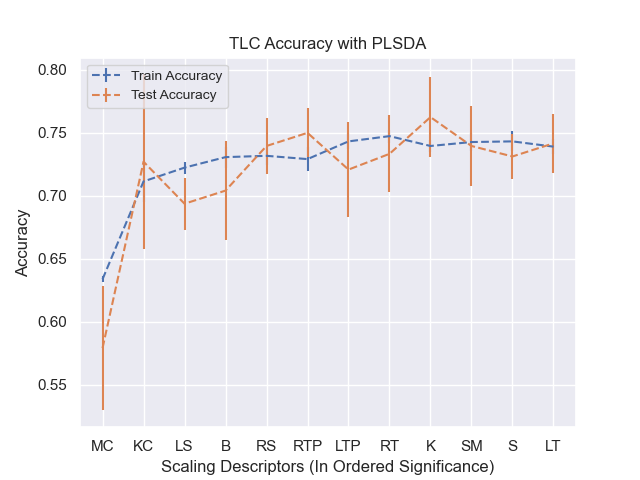}
  \caption{}
  \label{TLC_Scaling_Descriptor_Acc}
\end{subfigure}
\begin{subfigure}{.475\linewidth}
  \centering
  \includegraphics[width= 1\linewidth]{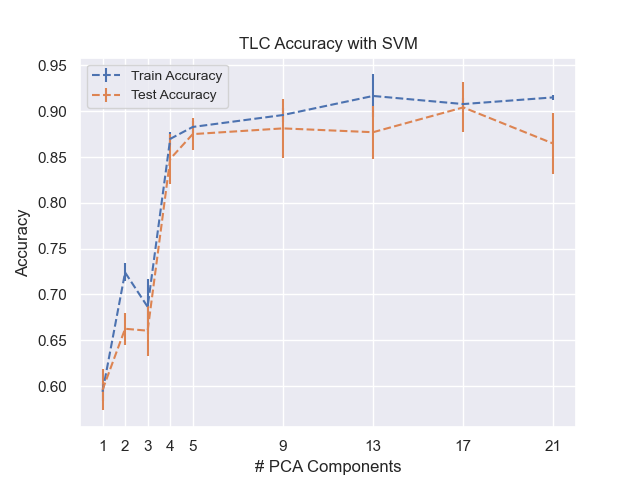}
  \caption{}
  \label{TLC_MIRS_Acc}
\end{subfigure}
\caption{Illustration of change in classifier performance (Training/Testing Accuracy) for total lactose content (TLC) with the introduction of (a)  scaling descriptors in order of 'significance' (b)  PCA Components.}
\label{fig-678}
\end{figure}

In building classifiers, Figure \ref{fig-678}, for instance,  depicts the change in classifier performance for Total Lactose Content (TLC) as the number of features escalates. The scaling descriptors were fed to the classifier in order of their importance (see in Table \ref{Best_descriptors}), namely mean curvature (MC), curvature central (KC), left slope (LS), broadness (B), right slope (RS), right tangent point (RTP), left tangent point (LTP), right tangent (RT), curvature (K), spectral mode (SM), slope (S), and left tangent (LT), as shown in Figure \ref{TLC_Scaling_Descriptor_Acc}. Optimal performance was achieved after including these descriptors, with no significant performance improvement observed upon adding slope (S) and left tangent (LT). The same methodology was applied to the PCA components, with the optimal number being 17 when only the PCA decomposed MIRS data was provided to the model.

Table \ref{Best Testing Accuracy} encapsulates the highest-performing classification model, along with the average testing accuracy and the number of features utilized for each milk quality parameter for both wavelet- and PCA-based methodologies. Additionally, classifier specifications (Table \ref{Classification_Method_Table}) are provided as supplementary information in the Appendix.

Generally, models based on scaling descriptors attain comparable performance to those based on PCA components, especially for technological and milk protein traits. However, models based on PCA components outperform those based on scaling descriptors when it comes to general milk quality traits. In terms of the number of discriminatory descriptors needed for optimal performance, models based on the scaling descriptors are relatively simpler than those based on PCA. Overall, scaling descriptors-based models achieve comparable performance with fewer predictors (half or less) than PCA-based models.

\begin{table*}[!t]
\centering
\caption{ Best performance summary of classification methods developed with the scaling descriptors and PCA components for each milk quality parameter investigated. The scaling descriptors used in the models can be found in Table \ref{Best_descriptors} in the Appendix. }
\begin{tabular}{|c|c|c|c|c|c|c|}
\hline
\multirow{3}{*}{ {\bf MQP} }& \multicolumn{3}{|c|}{ {\bf Scaling Descriptors} } & \multicolumn{3}{|c|}{ {\bf PCA components} } \\
\cline{2-7}
 & {\bf Model} & {\bf Acc.$\pm$Std.} &  {\bf \#Descriptors} & {\bf Model} & {\bf Acc.$\pm$Std.} &  {\bf \#PCA Components}\\
\hline
\multicolumn{7}{|c|}{ {\bf Technological traits} } \\
\hline
RCT & SVM   & 0.69$\pm$0.01 & 12  & Logit & 0.73$\pm$0.04 & 21\\
\hline
k20 & SVM & 0.70 $\pm$0.04 & 4 & SVM & 0.77 $\pm$0.01  & 5 \\
\hline
a30 & PLSDA & 0.71$\pm$0.02 & 9 & SVM  & 0.76$\pm$0.05 & 5 \\
\hline
a60 & SVM & 0.68 $\pm$0.03 & 3 & PLSDA    & 0.67$\pm$0.03 & 9 \\
\hline
CMS & Logit & 0.56$\pm$0.06 & 4 & SVM  & 0.63$\pm$0.03 & 17 \\
\hline
pH & SVM    & 0.68$\pm$0.03 & 10 & PLSDA & 0.83$\pm$0.02  & 21 \\
\hline
HS & SVM    & 0.71$\pm$0.02 & 10  & Logit & 0.74$\pm$0.02  & 21 \\
\hline
\multicolumn{7}{| c |}{ { \bf Milk Protein Traits} } \\
\hline
$\alpha_{S1}$-CN & PLSDA & 0.40$\pm$0.05 & 11 & SVM & 0.49$\pm$0.03 & 5 \\
\hline
$\alpha_{S2}$-CN & Logit   & 0.38$\pm$0.03 & 10 & LDA & 0.42$\pm$0.01 & 5 \\
\hline 
$\beta$-CN   & PLSDA & 0.35$\pm$0.04  & 8 & Logit & 0.51$\pm$0.05 & 21 \\
\hline 
$\kappa$-CN      & SVM & 0.41$\pm$0.03 & 5 & Logit & 0.49$\pm$0.03 & 9\\
\hline 
$\alpha$-L       & Logit & 0.37$\pm$0.03 & 10 & Logit & 0.42$\pm$0.03 & 13\\
\hline 
$\beta$-LGA      & LDA & 0.37$\pm$0.02 & 11 & GNB  & 0.42$\pm$0.03 & 13\\
\hline
$\beta$-LGB      & SVM   & 0.38$\pm$0.07 & 2  & Logit & 0.41$\pm$0.07 & 13\\
\hline
\multicolumn{7}{| c |}{ { \bf Milk Quality Traits} } \\
\hline
TLC               & Logit  & 0.74$\pm$0.04  & 6 & SVM & 0.90$\pm$0.03 & 17\\
\hline
TUC               & SVM & 0.76$\pm$0.04 & 7 & PLSDA & 0.90$\pm$0.03 & 13\\
\hline
TFC               & SVM & 0.67$\pm$0.01   & 7 & Logit & 0.94$\pm$0.01 & 4\\
\hline
TPC               & Logit & 0.45$\pm$0.02 & 10 & SVM & 0.81$\pm$0.01 & 13\\
\hline
\end{tabular}
\label{Best Testing Accuracy}
\end{table*}

\section{Discussion}\label{sec:Discussion}

This study demonstrates the scaling nature of milk spectral data and their importance in assessing milk quality. In the following, we discuss the general advantages of using scaling properties and their importance to the agri-community in particular. We also review some key challenges that lead to future studies.  

\subsection{General Advantages}
In this study, scaling descriptors have demonstrated promising performance when analyzing milk spectral data. The performance can be primarily assessed in two ways, namely, model simplicity and classification efficiency. 

Considering model simplicity,  models based on scaling descriptors are generally simpler in terms of the number of features as compared to models based on PCA components. For example, curd firmness at 60 minutes, a60, achieves 68\% accuracy with three scaling descriptors, while the PCA-based approach yields 67.0\% accuracy with 9 PCA components. Moreover, Frizzarin et al.\cite{Frizzarin2020} used wavelengths in milk spectra as discriminatory features to classify the same technological and protein traits. Comparing their model size with the classification performance listed in Table \ref{Best Testing Accuracy}, it is clear that scaling descriptors-based models are considerably simpler and achieve comparable performance for most MQPs.

The model simplicity and classification performance associated with the scaling descriptors can be attributed to their enhanced ability to capture complex dynamics in milk spectral data, thereby improving the capacity to develop statistical models with robust predictive capabilities \cite{Dixon2023, DixonMurmur2023}. In terms of wavelet-based strategies for analyzing milk spectral data, the wavelet transform offers unique benefits, particularly the ability to capture co-expression of certain components in the milk composition. Co-expression, where groups of milk components exhibit similar intensity patterns or correlated presence, may provide valuable information about their level of presence in milk samples. However, most existing methods like those by Frizzarin et al. \cite{Frizzarin2020} and Vimalajeewa et al. \cite{Dixon2018} use individual wave numbers and PCA components and overlook co-expression patterns. Moreover, given that pre-processing methods can vary, analysis outcomes may also differ depending on the methods used. The wavelet transform, requiring minimal pre-processing, guarantees minimal information loss advantages over existing methods. These factors underscore the potential of scaling descriptors in enhancing the analysis of milk spectral data.

In addition, traditional analyses of milk spectral data have often utilized partial least squares regression-based models, gaining prominence due to their consistently robust predictive performance \cite{Wold2001, Dixon2018}. However, as demonstrated in this study (see Table \ref{Best Testing Accuracy}), PLSR may only sometimes be the optimal choice. For instance, the study by \cite{Frizzarin2020} suggests that the PLSDA model is an optimal choice for technological MQPs, except for k20, a30, and CMS. In contrast, our wavelet- and PCA-based descriptors indicate that PLSDA is the best choice for a30, a60, and pH, respectively. Therefore, the most suitable model depends on several factors, including the type of features, milk quality parameters, and data pre-processing steps. This highlights the need for considering a pool of machine learning models when analyzing milk spectral data.

\subsection{Benefits to the Agri-Community}
Analyzing milk composition is crucial for enhancing milk-derived products and optimizing farming strategies. We will discuss these two key points and then explore how the proposed classification models could benefit them.

Cheese, a primary milk product, has its yield and quality determined by several factors. For example, attributes related to milk coagulation, such as increased curd firming capacity and reduced milk coagulation time, are positively associated with enhanced cheese yield \cite{Frizzarin2020}. Similarly, heat stability, CMS, and pH are essential for cheese production and other milk products. Technological traits in milk, such as $\alpha-S1$-CN, $\beta$-CN, $k$-CN, and $\beta$-LG, positively influence cheese production. Furthermore, modifications in milk content, along with changes in lactose, mineral content, and milk pH, lead to decreased cheese yield and altered manufacturing properties \cite{Grant2007}.

Examining the composition of milk solids can offer valuable insights for improving milk production processes. Milk solids comprise fat, protein, lactose, and minerals. Typical milk fat values range from 3.7\% (Holstein) to 4.9\% (Jersey); milk protein values range from 3.1\% (Holstein) to 3.8\% (Jersey); lactose levels are usually between 4.6\% and 4.8\% for all breeds, and minerals average 0.74\% \cite{Grant2007}. Deviations from these values provide useful information for producers. For example, maintaining milk fat and protein tests, heavily emphasized in milk pricing formulas, offers an economic advantage. The standard milk fat percentage also indicates cow health, while a low milk fat percentage may lead to sore feet, acidosis, and feed intake issues. Higher protein contributes to increased cheese yields. Conversely, milk protein depression occurs if the ratio of milk protein to milk fat falls below 0.8 for Holsteins. When this ratio exceeds 1, the herd is said to suffer from milk fat depression \cite{Grant2007}.

The proposed models can offer several ways to provide valuable insight into optimizing milk-related products and farming procedures. Most importantly, using these standard ranges instead of the quartile-based categorization used in this study could aid in automatically discriminating milk samples at the processing stage. For instance, higher protein composition is indicated by a greater multifractality than monofractality in  Table \ref{Scaling_descriptors_summaryStats_2}. On the other hand, the broadness of the total lactose content (TLC) and total fat content (TFC) in Table \ref{Scaling_descriptors_summaryStats_3} indicates that higher multifractality is associated with lower TLC and TFC levels. However, the situation is reversed when it comes to the amount of total urea content in the milk. 

Overall, these factors would enable the identification of milk samples that deviate from the milk composition standards defined by the authorities. Consequently, farmers could identify the corresponding cows promptly and take appropriate actions, such as maximizing feed intake, monitoring dietary composition, and properly feeding protein, energy, fiber, minerals, and vitamins \cite{Looper2001}.

\subsection{Challenges}
The proposed methodology presents challenges that need to be addressed in future studies. For example,  when assessing monofractality, the scale index range used to calculate slope values was restricted from 3 to 7, which may not be suitable for all MIR spectra and could result in slope values outside the theoretically expected range of -3 to -1. Similarly, selecting an appropriate range of moment, $q$, that is suitable for all spectra when calculating multifractal spectra also presented challenges. The range $q = [2, 9]$ applied to over 90\% of signals, but it resulted in numerical instability for some signals. As a result, a shorter range, $q = [4, 7]$ was used for which the range $q = [2, 9]$ did not work. Consequently, the generalizability of the proposed procedure is limited, and further investigation is needed to identify more generalizable values for these parameters. Additionally, the study used the discrete wavelet transform, which requires the signal length to be a power of two. Therefore, the maximum possible signal length that satisfies this condition was selected, resulting in some information loss. This can be overcome by replacing DWT with a non-decimated wavelet transform as it allows performing wavelet transforms on a signal of any length \cite{vidakovic1999}. 

In terms of comparing performance to previous studies, while the performance for the technological and protein traits vary between 62-80\% and 40-48\% in the study by Frizzarin et al.\cite{Frizzarin2020}, they range between 56-71\% and 35-40\% in the present study. This difference can be attributed as follows. In order to satisfy the requirement of some of the predictive models implemented, some of the data needed to be coerced into single observations. Where multiple observations of a particular cow were observed, we took the average (both the original study data and the produced scaling descriptors). However, the study by Frizzarin et al.\cite{Frizzarin2020} utilized the Mahalanobis distance to maximize the variability within their data. This could partially explain some of the differences observed between model performance. Overall, the lack of standardized data pre-processing steps limits the comparability of the model performance to those of existing studies.

Moreover, the scaling properties-based and PCA-based descriptors are extracted independently from two domains. It has been proved in the literature that the joint of such independently extracted features contributes to improved performance \cite{Wang2011, DixonOvarian2023}. Despite the increased number of features, the combined use of proposed descriptors with PCA-based features did not significantly improve performance. Hence, we did not include outcomes in this study. Nevertheless, it is interesting to investigate possible reasons why such joint feature approaches do not work. Therefore, this could be a potential direction for future research.

\section{Conclusions}\label{sec:Conclusions}
This study demonstrates the importance of accounting for scaling properties in milk spectral data for assessing milk quality. Scaling properties are assessed using monofractal and multifractal behaviors in the wavelet domain. The potential of accurately detecting a set of 18 different milk quality parameters (MQPs) of these features is assessed using eight commonly used classifiers. Applying proposed scaling descriptors on an MIRS milk quality dataset showed comparable performance for several MQPs. More specifically, compared to existing methods based on PCA, the proposed descriptors help build much simpler predictive models achieving comparable performance, particularly for technological and protein milk quality traits. 

Based on the findings of this study and related research in the literature, it is evident that there is still scope for enhancing performance. Integrating discriminatory features derived from independent domains has been demonstrated to be effective in other application areas \cite{Dixon2022}. This approach presents a potential avenue for continuing this study in the future. For example, combining frequency domain features (such as scaling descriptors) with wavelengths and PCA components may be promising.

This study provides a novel approach for milk quality assessment by incorporating the scaling properties of milk spectral data computed within the wavelet domain. This approach offers a fresh perspective compared to traditional methods that predominantly rely on the PCA of milk spectral data or wavenumbers corresponding to peak intensity values in the MIR spectra. Therefore, the scaling descriptors could serve as potential discriminatory features for automated milk quality assessment.

In the spirit of reproducible research, the software used in this paper is posted on \href{https://github.com/dmaywald/MIRS_Multifractal_Anaysis}{ {\it github repository}}.

\begin{acknowledgement}
The authors are grateful to Teagasc for providing the MIRS dataset and Dr. Brani Vidacovik, Texas A\&M university for providing Matlab Software for performing wavelet transforms. 
\end{acknowledgement}

\bibliographystyle{plain}
\bibliography{sample}

\begin{thebibliography}{10}

\bibitem{Molina2019}
Ana~Mar{\'\i}a Aguilar-Molina, Fernando Angulo-Brown, and Alejandro
  Mu{\~n}oz-Diosdado.
\newblock Multifractal spectrum curvature of rr tachograms of healthy people
  and patients with congestive heart failure, a new tool to assess health
  conditions.
\newblock {\em Entropy}, 21(6):581, 2019.

\bibitem{Alexandratos2012}
Nikos Alexandratos and Jelle Bruinsma.
\newblock World agriculture towards 2030/2050: the 2012 revision, 2012.

\bibitem{Amar2020}
H.~Amar, H.~Ghodbane, M.~Amir, M.~A. Zidane, C.~Hamouda, and A.~Rouane.
\newblock Microstrip sensor for product quality monitoring.
\newblock {\em Journal of Computational Electronics}, 19(3):1329--1336, 2020.

\bibitem{Frizzarin2020}
M~Frizzarin, I.~C. Gormely, D.~P. Berry, T.~B. Murphy, A~Casa, and A~Lynch.
\newblock Predicting cow milk quality traits from routinely available milk
  spectra using statistical machine learning methods.
\newblock {\em Dairy Science}, 104:7438--7447, 2020.

\bibitem{Goncalves1998}
P.~Goncalves, R.~Riedi, and R.~Baraniuk.
\newblock A simple statistical analysis of wavelet-based multifractal spectrum
  estimation.
\newblock In {\em Conference Record of Thirty-Second Asilomar Conference on
  Signals, Systems and Computers (Cat. No.98CH36284)}, volume~1, pages 287--291
  vol.1, 1998.

\bibitem{Grant2007}
Rick Grant and Paul~J. Kononoff.
\newblock Feeding to maximize milk protein and fat yields.
\newblock {\em Institute of Agriculture and Natural Resources}, 2007.

\bibitem{Dixon2023}
Jongphil Kim, Hin~Kyeol Woo, Dixon Vimalajeewa, and Brani Vidakovic.
\newblock Analysis and classification of 1h-nmr spectra by multifractal
  analysis.
\newblock {\em PLOS ONE}, 18(6):1--18, 06 2023.

\bibitem{Kunes2021}
Radim Kunes, Petr Bartos, Gustavo~Kenji Iwasaka, Ales Lang, Tomas Hankovec,
  Lubos Smutny, Pavel Cerny, Anna Poborska, Pavel Smetana, Pavel Kriz, and
  Nadezda Kernerova.
\newblock In-line technologies for the analysis of important milk parameters
  during the milking process: A review.
\newblock {\em Agriculture}, 11(3), 2021.

\bibitem{Looper2001}
Michael Looper, Sandra~R. Stokes, Dan~N. Waldner, and Ellen~R. Jordan.
\newblock Managing milk composition: Evaluating herd potential.
\newblock {\em College of Agriculture, Consumer and Environmental Sciences New
  Mexico State University}, 2001.

\bibitem{Mallat1989}
S.G. Mallat.
\newblock A theory for multiresolution signal decomposition: the wavelet
  representation.
\newblock {\em IEEE Transactions on Pattern Analysis and Machine Intelligence},
  11(7):674--693, 1989.

\bibitem{Su2019}
Wen-Hao Su and Da-Wen Sun.
\newblock Mid-infrared (mir) spectroscopy for quality analysis of liquid foods.
\newblock {\em Food Engineering Reviews}, 11(3):142--158, 2019.

\bibitem{Helan2023}
Helan~Vidhya T, Sarveswaran S, Sejal Jha, and Soundarya B.
\newblock Milksafe: A hardware-enabled milk quality prediction using machine
  learning.
\newblock In {\em 2023 2nd International Conference on Vision Towards Emerging
  Trends in Communication and Networking Technologies (ViTECoN)}, pages 1--6,
  2023.

\bibitem{Tsenkova2022}
Roumiana Tsenkova and Jelena Muncan.
\newblock {\em Wavelet Transform of Near-Infrared Individual Cow's Milk for
  Single-Spectrum Mastitis Diagnosis}, pages 289--296.
\newblock Springer Singapore, Singapore, 2022.

\bibitem{vidakovic1999}
B~Vidakovic.
\newblock Statistical modeling by wavelets. {J}ohn {W}iley and {S}ons, 1999.

\bibitem{Vidhya2023}
S~Vidhya, V~Siva Vadivu~Ragavi, JK~Monica, and B~Kanisha.
\newblock Milk quality prediction using supervised machine learning technique.
\newblock In {\em International Conference on Emerging Trends in Expert
  Applications \& Security}, pages 267--278. Springer, 2023.

\bibitem{Dixon2017}
Dixon Vimalajeewa, Donagh Berry, Eric Robson, and Chamil Kulatunga.
\newblock Evaluation of non-linearity in mir spectroscopic data for compressed
  learning.
\newblock In {\em 2017 IEEE International Conference on Data Mining Workshops
  (ICDMW)}, pages 545--552, 2017.

\bibitem{DixonOvarian2023}
Dixon Vimalajeewa, Scott~Alan Bruce, and Brani Vidakovic.
\newblock Early detection of ovarian cancer by wavelet analysis of protein mass
  spectra.
\newblock {\em Statistics in Medicine}, 42(13):2257--2273, 2023.

\bibitem{Dixon2018}
Dixon Vimalajeewa, Chamil Kulatunga, and Donagh~P. Berry.
\newblock Learning in the compressed data domain: Application to milk quality
  prediction.
\newblock {\em Information Sciences}, 459:149--167, 2018.

\bibitem{Dixon2022a}
Dixon Vimalajeewa, Chamil Kulatunga, Donagh~P. Berry, and Sasitharan
  Balasubramaniam.
\newblock A service-based joint model used for distributed learning:
  Application for smart agriculture.
\newblock {\em IEEE Transactions on Emerging Topics in Computing},
  10(2):838--854, 2022.

\bibitem{DixonMurmur2023}
Dixon Vimalajeewa, Chihoon Lee, and Brani Vidakovic.
\newblock A method for detecting murmurous heart sounds based on self-similar
  properties.
\newblock {\em arXiv preprint arXiv:2306.05283}, 2023.

\bibitem{Dixon2022}
Dixon Vimalajeewa, Ethan McDonald, Scott~Alan Bruce, and Brani Vidakovic.
\newblock Wavelet-based approach for diagnosing attention deficit hyperactivity
  disorder (adhd).
\newblock {\em Scientific Reports}, 12(1):21928, 2022.

\bibitem{Dixon2023c}
Dixon Vimalajeewa, Ethan McDonald, Megan Tung, and Brani Vidakovic.
\newblock Parkinson’s disease diagnosis with gait characteristics extracted
  using wavelet transforms.
\newblock {\em IEEE Journal of Translational Engineering in Health and
  Medicine}, 11:271--281, 2023.

\bibitem{Wang2011}
Fei Wang, Xian-Hua Han, and Yen-Wei Chen.
\newblock Biomedical imaging modality classification using combined visual
  features and textual terms.
\newblock {\em International Journal of Biomedical Imaging}, 2011:241396, 2011.

\bibitem{Wold2001}
Svante Wold, Michael Sj{\"o}str{\"o}m, and Lennart Eriksson.
\newblock Pls-regression: a basic tool of chemometrics.
\newblock {\em Chemometrics and intelligent laboratory systems},
  58(2):109--130, 2001.

\end{thebibliography}
\newpage
\section*{Appendix}
\addcontentsline{toc}{section}{Appendix}
\begin{table*}[!h]
\centering
\caption{The Classification Methods explored in the present study along with Tuning Parameter Considered}
\begin{tabular}{|c|l|l|}
\hline
{\bf } & {\bf Model} & {\bf Model Parameters } \\
\hline
\multirow{3}{*}{1}  & \multirow{3}{*}{Support Vector Machine (SVM)}  &  C: l2 Regularization \\ && kernel: kernel type used (Linear or RBF)\\ && gamma: Kernel coefficient for ‘RBF’ \\
\hline
\multirow{2}{*}{2} & \multirow{2}{*}{Linear Discriminant Analysis (LDA)} & shrinkage: regularization of covariance \\&& matrices of the classes\\
\hline
\multirow{5}{*}{3} & \multirow{5}{150pt}{Gradient Boosting Machine (GBM)} & max depth: Max depth of estimators\\
&& learning rate: Contribution of trees\\
&& n estimators: Number of boosting stages\\
&& subsample: fraction of samples used when \\&& 
fitting individual base learners\\
\hline
\multirow{2}{*}{4} &  \multirow{2}{*}{K-Nearest Neighbors (KNN)}  & 
n neighbors: Number of neighbors used \\
&& p: lp distance used (l1 or l2) \\

\hline
\multirow{3}{*}{5} &  \multirow{3}{*}{Logistic Regression (Logit)}  & 
1l ratio: Ratio of l1 and l2 regularization \\ 
&& in elastic-net penalization\\
&& C: Regularization Strength \\
\hline
\multirow{2}{*}{6} & \multirow{2}{*}{Gaussian Naive Bayes (GNB)} &  \multirow{2}{*}{-}\\&& \\
\hline
\multirow{2}{*}{7} & \multirow{2}{150pt}{Quadratic Discriminant Analysis (QDA)} &  \multirow{2}{*}{-}\\&& \\
\hline
\multirow{2}{*}{8} & \multirow{2}{175pt}{Partial Least Squares Discriminant Analysis (PLSDA)} &  
n components: Number of latent variables \\&& kept in transformation\\
\hline

\end{tabular}
\label{Classification_Method_Table}
\end{table*}

\newpage
\restoregeometry
\newgeometry{bottom=10mm}
\begin{landscape}
\begin{table}[h]
\centering
\caption{Variability ($mean \pm std$) of scaling descriptors for the technological traits.}
\begin{tabular}{|c|c|c|c|c|c|c|c|c|c|c|c|c|c|}
\hline
\multirow{2}{*}{\bf MQP} & 
\multirow{2}{*}{\bf Category} &
\multicolumn{12}{|c|}{ {\bf Scaling Descriptors} }\\
\cline {3-14}
 & & {\bf S} &{\bf B} &{\bf SM} &{\bf MC} &{\bf K} &{\bf KC} &{\bf LS} &{\bf RS} &{\bf LT} &{\bf RT} &{\bf LTP} &{\bf RTP} \\
\hline
\hline
\multirow{2}{*}{RCT} & Q Low & 
-1.886 $\pm$ 0.17 & 
1.975 $\pm$ 0.17 & 
1.497 $\pm$ 0.06 & 
0.356 $\pm$ 0.06 & 
0.0191 $\pm$ 0.002 & 
0.027 $\pm$ 0.0 & 
0.186 $\pm$ 0.01 & 
-0.23 $\pm$ 0.04 & 
0.412 $\pm$ 0.03 & 
-0.612 $\pm$ 0.14 & 
0.416 $\pm$ 0.06 & 
2.391 $\pm$ 0.15 \\
\cline{2-14} & Q High & 
-1.873 $\pm$ 0.19 & 
1.864 $\pm$ 0.19 & 
1.452 $\pm$ 0.07 & 
0.398 $\pm$ 0.07 & 
0.0184 $\pm$ 0.002 & 
0.025 $\pm$ 0.0 & 
0.193 $\pm$ 0.01 & 
-0.252 $\pm$ 0.07 & 
0.413 $\pm$ 0.03 & 
-0.66 $\pm$ 0.18 & 
0.413 $\pm$ 0.07 & 
2.277 $\pm$ 0.18 \\
\hline
\hline
\multirow{2}{*}{k20} & Q Low & 
-1.901 $\pm$ 0.18 & 
1.97 $\pm$ 0.2 & 
1.497 $\pm$ 0.06 & 
0.358 $\pm$ 0.07 & 
0.0189 $\pm$ 0.002 & 
0.027 $\pm$ 0.0 & 
0.187 $\pm$ 0.02 & 
-0.233 $\pm$ 0.06 & 
0.415 $\pm$ 0.03 & 
-0.617 $\pm$ 0.17 & 
0.418 $\pm$ 0.07 & 
2.388 $\pm$ 0.17 \\
\cline{2-14} & Q High & 
-1.864 $\pm$ 0.18 & 
1.882 $\pm$ 0.16 & 
1.457 $\pm$ 0.07 & 
0.392 $\pm$ 0.06 & 
0.0186 $\pm$ 0.002 & 
0.025 $\pm$ 0.0 & 
0.192 $\pm$ 0.01 & 
-0.245 $\pm$ 0.04 & 
0.411 $\pm$ 0.03 & 
-0.648 $\pm$ 0.16 & 
0.412 $\pm$ 0.07 & 
2.294 $\pm$ 0.15 \\
\hline
\hline
\multirow{2}{*}{a30} & Q Low & 
-1.879 $\pm$ 0.19 & 
1.87 $\pm$ 0.17 & 
1.456 $\pm$ 0.07 & 
0.395 $\pm$ 0.06 & 
0.0185 $\pm$ 0.002 & 
0.025 $\pm$ 0.0 & 
0.193 $\pm$ 0.01 & 
-0.251 $\pm$ 0.07 & 
0.414 $\pm$ 0.03 & 
-0.655 $\pm$ 0.18 & 
0.416 $\pm$ 0.07 & 
2.285 $\pm$ 0.16 \\
\cline{2-14} & Q High & 
-1.895 $\pm$ 0.17 & 
1.992 $\pm$ 0.18 & 
1.503 $\pm$ 0.06 & 
0.351 $\pm$ 0.07 & 
0.0191 $\pm$ 0.002 & 
0.028 $\pm$ 0.0 & 
0.185 $\pm$ 0.01 & 
-0.225 $\pm$ 0.03 & 
0.412 $\pm$ 0.03 & 
-0.602 $\pm$ 0.14 & 
0.417 $\pm$ 0.07 & 
2.409 $\pm$ 0.15 \\
\hline
\hline
\multirow{2}{*}{a30} & Q Low & 
-1.898 $\pm$ 0.19 & 
1.861 $\pm$ 0.17 & 
1.464 $\pm$ 0.07 & 
0.398 $\pm$ 0.06 & 
0.0184 $\pm$ 0.002 & 
0.025 $\pm$ 0.0 & 
0.194 $\pm$ 0.01 & 
-0.252 $\pm$ 0.06 & 
0.414 $\pm$ 0.03 & 
-0.66 $\pm$ 0.16 & 
0.427 $\pm$ 0.07 & 
2.288 $\pm$ 0.17 \\
\cline{2-14} & Q High & 
-1.862 $\pm$ 0.17 & 
1.972 $\pm$ 0.18 & 
1.483 $\pm$ 0.07 & 
0.358 $\pm$ 0.07 & 
0.0191 $\pm$ 0.002 & 
0.027 $\pm$ 0.0 & 
0.186 $\pm$ 0.01 & 
-0.231 $\pm$ 0.06 & 
0.41 $\pm$ 0.03 & 
-0.615 $\pm$ 0.16 & 
0.403 $\pm$ 0.07 & 
2.375 $\pm$ 0.17 \\
\hline
\hline
\multirow{2}{*}{CMS} & Q Low & 
-1.905 $\pm$ 0.2 & 
1.904 $\pm$ 0.21 & 
1.475 $\pm$ 0.07 & 
0.386 $\pm$ 0.1 & 
0.0186 $\pm$ 0.002 & 
0.026 $\pm$ 0.0 & 
0.191 $\pm$ 0.02 & 
-0.252 $\pm$ 0.14 & 
0.419 $\pm$ 0.04 & 
-0.642 $\pm$ 0.19 & 
0.423 $\pm$ 0.07 & 
2.327 $\pm$ 0.19 \\
\cline{2-14} & Q High & 
-1.876 $\pm$ 0.19 & 
1.935 $\pm$ 0.19 & 
1.477 $\pm$ 0.07 & 
0.372 $\pm$ 0.07 & 
0.0188 $\pm$ 0.002 & 
0.026 $\pm$ 0.0 & 
0.189 $\pm$ 0.01 & 
-0.236 $\pm$ 0.04 & 
0.413 $\pm$ 0.03 & 
-0.635 $\pm$ 0.15 & 
0.411 $\pm$ 0.07 & 
2.345 $\pm$ 0.17 \\
\hline
\hline
\multirow{2}{*}{pH} & Q Low & 
-1.91 $\pm$ 0.2 & 
1.938 $\pm$ 0.21 & 
1.484 $\pm$ 0.06 & 
0.373 $\pm$ 0.1 & 
0.0186 $\pm$ 0.002 & 
0.027 $\pm$ 0.0 & 
0.189 $\pm$ 0.02 & 
-0.243 $\pm$ 0.12 & 
0.419 $\pm$ 0.04 & 
-0.634 $\pm$ 0.16 & 
0.419 $\pm$ 0.07 & 
2.357 $\pm$ 0.17 \\
\cline{2-14} & Q High & 
-1.861 $\pm$ 0.19 & 
1.905 $\pm$ 0.19 & 
1.467 $\pm$ 0.08 & 
0.383 $\pm$ 0.07 & 
0.0189 $\pm$ 0.002 & 
0.026 $\pm$ 0.0 & 
0.191 $\pm$ 0.01 & 
-0.245 $\pm$ 0.07 & 
0.411 $\pm$ 0.03 & 
-0.641 $\pm$ 0.17 & 
0.412 $\pm$ 0.07 & 
2.317 $\pm$ 0.19 \\
\hline
\hline 
\multirow{2}{*}{HS} & Q Low & 
-1.914 $\pm$ 0.2 & 
1.831 $\pm$ 0.15 & 
1.456 $\pm$ 0.07 & 
0.412 $\pm$ 0.05 & 
0.0181 $\pm$ 0.002 & 
0.024 $\pm$ 0.0 & 
0.197 $\pm$ 0.01 & 
-0.253 $\pm$ 0.05 & 
0.417 $\pm$ 0.04 & 
-0.656 $\pm$ 0.15 & 
0.434 $\pm$ 0.07 & 
2.265 $\pm$ 0.14 \\
\cline{2-14} & Q High & 
-1.843 $\pm$ 0.17 & 
1.958 $\pm$ 0.19 & 
1.485 $\pm$ 0.07 & 
0.362 $\pm$ 0.07 & 
0.0192 $\pm$ 0.002 & 
0.027 $\pm$ 0.0 & 
0.186 $\pm$ 0.01 & 
-0.237 $\pm$ 0.07 & 
0.409 $\pm$ 0.03 & 
-0.637 $\pm$ 0.18 & 
0.405 $\pm$ 0.06 & 
2.363 $\pm$ 0.18 \\
\hline

\end{tabular}
\label{Scaling_descriptors_summaryStats_1}
\end{table}
\end{landscape}
\restoregeometry

\newpage
\restoregeometry
\newgeometry{bottom=10mm}
\begin{landscape}
\begin{table}[h]
\centering
\caption{Variability ($mean \pm std$) of scaling descriptors for the milk protein traits.}
\begin{tabular}{|c|c|c|c|c|c|c|c|c|c|c|c|c|c|}
\hline
\multirow{2}{*}{\bf MQP} & 
\multirow{2}{*}{\bf Category} &
\multicolumn{12}{|c|}{ {\bf Scaling Descriptors} }\\
\cline {3-14}
 & & {\bf S} &{\bf B} &{\bf SM} &{\bf MC} &{\bf K} &{\bf KC} &{\bf LS} &{\bf RS} &{\bf LT} &{\bf RT} &{\bf LTP} &{\bf RTP} \\
\hline
\hline
\multirow{4}{*}{$\alpha_{\text{S1}}$-CN} & Q1 & 
-1.886 $\pm$ 0.18 & 
1.875 $\pm$ 0.2 & 
1.463 $\pm$ 0.06 & 
0.402 $\pm$ 0.13 & 
0.0186 $\pm$ 0.002 & 
0.025 $\pm$ 0.0 & 
0.193 $\pm$ 0.01 & 
-0.263 $\pm$ 0.18 & 
0.415 $\pm$ 0.03 & 
-0.656 $\pm$ 0.18 & 
0.42 $\pm$ 0.07 & 
2.295 $\pm$ 0.18 \\
\cline{2-14} & Q2 & 
-1.897 $\pm$ 0.19 & 
1.93 $\pm$ 0.2 & 
1.48 $\pm$ 0.05 & 
0.375 $\pm$ 0.07 & 
0.0188 $\pm$ 0.003 & 
0.026 $\pm$ 0.0 & 
0.19 $\pm$ 0.02 & 
-0.239 $\pm$ 0.05 & 
0.416 $\pm$ 0.04 & 
-0.617 $\pm$ 0.14 & 
0.419 $\pm$ 0.06 & 
2.35 $\pm$ 0.17 \\
\cline{2-14} & Q3 & 
-1.921 $\pm$ 0.21 & 
1.926 $\pm$ 0.22 & 
1.483 $\pm$ 0.06 & 
0.377 $\pm$ 0.08 & 
0.0185 $\pm$ 0.002 & 
0.026 $\pm$ 0.01 & 
0.191 $\pm$ 0.02 & 
-0.24 $\pm$ 0.07 & 
0.423 $\pm$ 0.04 & 
-0.644 $\pm$ 0.17 & 
0.425 $\pm$ 0.08 & 
2.351 $\pm$ 0.17 \\
\cline{2-14} & Q4 & 
-1.95 $\pm$ 0.21 & 
1.961 $\pm$ 0.2 & 
1.501 $\pm$ 0.05 & 
0.36 $\pm$ 0.07 & 
0.0185 $\pm$ 0.002 & 
0.027 $\pm$ 0.0 & 
0.188 $\pm$ 0.02 & 
-0.23 $\pm$ 0.04 & 
0.422 $\pm$ 0.04 & 
-0.627 $\pm$ 0.16 & 
0.429 $\pm$ 0.08 & 
2.39 $\pm$ 0.16 \\
\hline
\hline
\multirow{4}{*}{$\alpha_{\text{S2}}$-CN} & Q1 & 
-1.859 $\pm$ 0.18 & 
1.929 $\pm$ 0.2 & 
1.48 $\pm$ 0.07 & 
0.373 $\pm$ 0.07 & 
0.019 $\pm$ 0.002 & 
0.026 $\pm$ 0.0 & 
0.188 $\pm$ 0.01 & 
-0.242 $\pm$ 0.06 & 
0.409 $\pm$ 0.03 & 
-0.639 $\pm$ 0.17 & 
0.413 $\pm$ 0.06 & 
2.342 $\pm$ 0.18 \\
\cline{2-14} & Q2 & 
-1.914 $\pm$ 0.2 & 
1.898 $\pm$ 0.22 & 
1.47 $\pm$ 0.06 & 
0.393 $\pm$ 0.13 & 
0.0185 $\pm$ 0.002 & 
0.026 $\pm$ 0.0 & 
0.192 $\pm$ 0.02 & 
-0.26 $\pm$ 0.19 & 
0.42 $\pm$ 0.04 & 
-0.642 $\pm$ 0.17 & 
0.421 $\pm$ 0.08 & 
2.319 $\pm$ 0.19 \\
\cline{2-14} & Q3 & 
-1.963 $\pm$ 0.2 & 
1.875 $\pm$ 0.2 & 
1.474 $\pm$ 0.05 & 
0.394 $\pm$ 0.08 & 
0.0178 $\pm$ 0.002 & 
0.025 $\pm$ 0.0 & 
0.195 $\pm$ 0.02 & 
-0.245 $\pm$ 0.04 & 
0.43 $\pm$ 0.04 & 
-0.672 $\pm$ 0.18 & 
0.437 $\pm$ 0.07 & 
2.313 $\pm$ 0.15 \\
\cline{2-14} & Q4 & 
-1.914 $\pm$ 0.2 & 
1.991 $\pm$ 0.2 & 
1.503 $\pm$ 0.06 & 
0.354 $\pm$ 0.07 & 
0.0191 $\pm$ 0.002 & 
0.028 $\pm$ 0.0 & 
0.186 $\pm$ 0.02 & 
-0.225 $\pm$ 0.04 & 
0.417 $\pm$ 0.04 & 
-0.592 $\pm$ 0.12 & 
0.421 $\pm$ 0.07 & 
2.412 $\pm$ 0.16 \\
\hline
\hline
\multirow{4}{*}{$\beta$-CN} & Q1 & 
-1.89 $\pm$ 0.18 & 
1.889 $\pm$ 0.22 & 
1.473 $\pm$ 0.06 & 
0.392 $\pm$ 0.13 & 
0.0186 $\pm$ 0.003 & 
0.025 $\pm$ 0.0 & 
0.191 $\pm$ 0.02 & 
-0.266 $\pm$ 0.19 & 
0.413 $\pm$ 0.03 & 
-0.662 $\pm$ 0.19 & 
0.42 $\pm$ 0.07 & 
2.309 $\pm$ 0.2 \\
\cline{2-14} & Q2 & 
-1.881 $\pm$ 0.2 & 
1.932 $\pm$ 0.19 & 
1.482 $\pm$ 0.06 & 
0.373 $\pm$ 0.07 & 
0.0189 $\pm$ 0.002 & 
0.026 $\pm$ 0.0 & 
0.189 $\pm$ 0.01 & 
-0.237 $\pm$ 0.05 & 
0.414 $\pm$ 0.03 & 
-0.63 $\pm$ 0.17 & 
0.417 $\pm$ 0.07 & 
2.349 $\pm$ 0.16 \\
\cline{2-14} & Q3 & 
-1.933 $\pm$ 0.21 & 
1.93 $\pm$ 0.19 & 
1.483 $\pm$ 0.06 & 
0.377 $\pm$ 0.08 & 
0.0185 $\pm$ 0.002 & 
0.026 $\pm$ 0.0 & 
0.191 $\pm$ 0.02 & 
-0.234 $\pm$ 0.04 & 
0.424 $\pm$ 0.04 & 
-0.627 $\pm$ 0.14 & 
0.427 $\pm$ 0.07 & 
2.358 $\pm$ 0.15 \\
\cline{2-14} & Q4 & 
-1.948 $\pm$ 0.2 & 
1.948 $\pm$ 0.22 & 
1.492 $\pm$ 0.06 & 
0.369 $\pm$ 0.08 & 
0.0185 $\pm$ 0.003 & 
0.027 $\pm$ 0.01 & 
0.19 $\pm$ 0.02 & 
-0.232 $\pm$ 0.04 & 
0.425 $\pm$ 0.04 & 
-0.62 $\pm$ 0.16 & 
0.429 $\pm$ 0.07 & 
2.377 $\pm$ 0.17 \\
\hline
\hline

\multirow{4}{*}{$\kappa$-CN} & Q1 & 
-1.919 $\pm$ 0.19 & 
1.846 $\pm$ 0.18 & 
1.462 $\pm$ 0.06 & 
0.405 $\pm$ 0.07 & 
0.018 $\pm$ 0.002 & 
0.024 $\pm$ 0.0 & 
0.195 $\pm$ 0.01 & 
-0.252 $\pm$ 0.04 & 
0.421 $\pm$ 0.04 & 
-0.675 $\pm$ 0.17 & 
0.432 $\pm$ 0.06 & 
2.278 $\pm$ 0.16 \\
\cline{2-14} & Q2 & 
-1.911 $\pm$ 0.21 & 
1.893 $\pm$ 0.22 & 
1.472 $\pm$ 0.06 & 
0.393 $\pm$ 0.13 & 
0.0184 $\pm$ 0.003 & 
0.026 $\pm$ 0.0 & 
0.192 $\pm$ 0.02 & 
-0.262 $\pm$ 0.19 & 
0.419 $\pm$ 0.04 & 
-0.661 $\pm$ 0.2 & 
0.423 $\pm$ 0.08 & 
2.316 $\pm$ 0.19 \\
\cline{2-14} & Q3 & 
-1.929 $\pm$ 0.23 & 
1.927 $\pm$ 0.21 & 
1.483 $\pm$ 0.06 & 
0.377 $\pm$ 0.08 & 
0.0186 $\pm$ 0.003 & 
0.026 $\pm$ 0.0 & 
0.191 $\pm$ 0.02 & 
-0.239 $\pm$ 0.07 & 
0.425 $\pm$ 0.04 & 
-0.625 $\pm$ 0.15 & 
0.427 $\pm$ 0.08 & 
2.353 $\pm$ 0.17 \\
\cline{2-14} & Q4 & 
-1.894 $\pm$ 0.18 & 
2.023 $\pm$ 0.17 & 
1.507 $\pm$ 0.05 & 
0.34 $\pm$ 0.06 & 
0.0192 $\pm$ 0.002 & 
0.028 $\pm$ 0.0 & 
0.183 $\pm$ 0.01 & 
-0.22 $\pm$ 0.03 & 
0.412 $\pm$ 0.03 & 
-0.586 $\pm$ 0.12 & 
0.41 $\pm$ 0.07 & 
2.433 $\pm$ 0.14 \\
\hline
\hline

\multirow{4}{*}{$\alpha$-L} & Q1 & 
-1.879 $\pm$ 0.17 & 
1.941 $\pm$ 0.18 & 
1.474 $\pm$ 0.06 & 
0.369 $\pm$ 0.07 & 
0.0188 $\pm$ 0.002 & 
0.026 $\pm$ 0.0 & 
0.188 $\pm$ 0.01 & 
-0.234 $\pm$ 0.04 & 
0.41 $\pm$ 0.03 & 
-0.627 $\pm$ 0.16 & 
0.407 $\pm$ 0.07 & 
2.348 $\pm$ 0.16 \\
\cline{2-14} & Q2 & 
-1.889 $\pm$ 0.2 & 
1.921 $\pm$ 0.19 & 
1.484 $\pm$ 0.05 & 
0.376 $\pm$ 0.07 & 
0.0187 $\pm$ 0.002 & 
0.026 $\pm$ 0.0 & 
0.19 $\pm$ 0.01 & 
-0.241 $\pm$ 0.05 & 
0.414 $\pm$ 0.03 & 
-0.635 $\pm$ 0.16 & 
0.423 $\pm$ 0.08 & 
2.344 $\pm$ 0.15 \\
\cline{2-14} & Q3 & 
-1.927 $\pm$ 0.2 & 
1.905 $\pm$ 0.22 & 
1.477 $\pm$ 0.06 & 
0.391 $\pm$ 0.13 & 
0.0185 $\pm$ 0.002 & 
0.026 $\pm$ 0.0 & 
0.192 $\pm$ 0.02 & 
-0.256 $\pm$ 0.18 & 
0.423 $\pm$ 0.04 & 
-0.645 $\pm$ 0.18 & 
0.427 $\pm$ 0.07 & 
2.332 $\pm$ 0.19 \\
\cline{2-14} & Q4 & 
-1.954 $\pm$ 0.22 & 
1.931 $\pm$ 0.24 & 
1.493 $\pm$ 0.07 & 
0.376 $\pm$ 0.09 & 
0.0184 $\pm$ 0.003 & 
0.026 $\pm$ 0.01 & 
0.191 $\pm$ 0.02 & 
-0.24 $\pm$ 0.07 & 
0.428 $\pm$ 0.04 & 
-0.634 $\pm$ 0.17 & 
0.435 $\pm$ 0.07 & 
2.367 $\pm$ 0.2 \\
\hline
\hline

\multirow{4}{*}{$\beta$-LGA} & Q1 & 
-1.913 $\pm$ 0.19 & 
1.885 $\pm$ 0.19 & 
1.468 $\pm$ 0.06 & 
0.393 $\pm$ 0.07 & 
0.0184 $\pm$ 0.002 & 
0.025 $\pm$ 0.0 & 
0.194 $\pm$ 0.02 & 
-0.244 $\pm$ 0.05 & 
0.418 $\pm$ 0.04 & 
-0.637 $\pm$ 0.15 & 
0.428 $\pm$ 0.07 & 
2.313 $\pm$ 0.16 \\
\cline{2-14} & Q2 & 
-1.936 $\pm$ 0.21 & 
1.882 $\pm$ 0.2 & 
1.478 $\pm$ 0.06 & 
0.389 $\pm$ 0.07 & 
0.0182 $\pm$ 0.002 & 
0.025 $\pm$ 0.0 & 
0.193 $\pm$ 0.02 & 
-0.249 $\pm$ 0.07 & 
0.422 $\pm$ 0.04 & 
-0.67 $\pm$ 0.19 & 
0.434 $\pm$ 0.07 & 
2.316 $\pm$ 0.17 \\
\cline{2-14} & Q3 & 
-1.905 $\pm$ 0.2 & 
1.935 $\pm$ 0.2 & 
1.482 $\pm$ 0.06 & 
0.373 $\pm$ 0.07 & 
0.0187 $\pm$ 0.002 & 
0.026 $\pm$ 0.0 & 
0.19 $\pm$ 0.02 & 
-0.237 $\pm$ 0.05 & 
0.418 $\pm$ 0.04 & 
-0.623 $\pm$ 0.13 & 
0.419 $\pm$ 0.07 & 
2.355 $\pm$ 0.16 \\
\cline{2-14} & Q4 & 
-1.903 $\pm$ 0.2 & 
1.991 $\pm$ 0.23 & 
1.5 $\pm$ 0.06 & 
0.359 $\pm$ 0.14 & 
0.0191 $\pm$ 0.003 & 
0.028 $\pm$ 0.01 & 
0.186 $\pm$ 0.02 & 
-0.242 $\pm$ 0.18 & 
0.418 $\pm$ 0.04 & 
-0.613 $\pm$ 0.18 & 
0.413 $\pm$ 0.07 & 
2.403 $\pm$ 0.2 \\
\hline
\hline

\multirow{4}{*}{$\beta$-LGB} & Q1 & 
-1.952 $\pm$ 0.2 & 
1.899 $\pm$ 0.24 & 
1.482 $\pm$ 0.07 & 
0.399 $\pm$ 0.14 & 
0.0185 $\pm$ 0.003 & 
0.025 $\pm$ 0.01 & 
0.193 $\pm$ 0.02 & 
-0.257 $\pm$ 0.19 & 
0.423 $\pm$ 0.04 & 
-0.628 $\pm$ 0.16 & 
0.436 $\pm$ 0.07 & 
2.336 $\pm$ 0.2 \\
\cline{2-14} & Q2 & 
-1.893 $\pm$ 0.17 & 
1.879 $\pm$ 0.2 & 
1.466 $\pm$ 0.05 & 
0.392 $\pm$ 0.07 & 
0.0184 $\pm$ 0.002 & 
0.025 $\pm$ 0.0 & 
0.193 $\pm$ 0.01 & 
-0.252 $\pm$ 0.08 & 
0.414 $\pm$ 0.03 & 
-0.655 $\pm$ 0.18 & 
0.422 $\pm$ 0.06 & 
2.301 $\pm$ 0.17 \\
\cline{2-14} & Q3 & 
-1.922 $\pm$ 0.23 & 
1.933 $\pm$ 0.18 & 
1.486 $\pm$ 0.05 & 
0.372 $\pm$ 0.07 & 
0.0185 $\pm$ 0.002 & 
0.026 $\pm$ 0.0 & 
0.19 $\pm$ 0.02 & 
-0.234 $\pm$ 0.03 & 
0.423 $\pm$ 0.04 & 
-0.641 $\pm$ 0.16 & 
0.425 $\pm$ 0.08 & 
2.357 $\pm$ 0.14 \\
\cline{2-14} & Q4 & 
-1.906 $\pm$ 0.2 & 
1.975 $\pm$ 0.21 & 
1.495 $\pm$ 0.07 & 
0.357 $\pm$ 0.08 & 
0.0188 $\pm$ 0.002 & 
0.027 $\pm$ 0.0 & 
0.187 $\pm$ 0.02 & 
-0.231 $\pm$ 0.05 & 
0.418 $\pm$ 0.04 & 
-0.611 $\pm$ 0.14 & 
0.417 $\pm$ 0.07 & 
2.392 $\pm$ 0.17 \\
\hline

\end{tabular}
\label{Scaling_descriptors_summaryStats_2}
\end{table}
\end{landscape}
\restoregeometry

\newpage
\restoregeometry
\newgeometry{bottom=10mm}
\begin{landscape}
\begin{table}[h]
\centering
\caption{Variability ($mean \pm std$) of scaling descriptors for the general milk quality traits.}
\begin{tabular}{|c|c|c|c|c|c|c|c|c|c|c|c|c|c|}
\hline
\multirow{2}{*}{\bf MQP} & 
\multirow{2}{*}{\bf Category} &
\multicolumn{12}{|c|}{ {\bf Scaling Descriptors} }\\
\cline {3-14}
 & & {\bf S} &{\bf B} &{\bf SM} &{\bf MC} &{\bf K} &{\bf KC} &{\bf LS} &{\bf RS} &{\bf LT} &{\bf RT} &{\bf LTP} &{\bf RTP} \\
\hline
\multicolumn{14}{|c|}{ {\bf Milk Quality Traits} } \\
\hline
\hline

\multirow{2}{*}{TLC} & Q Low & 
-1.881 $\pm$ 0.2 & 
1.979 $\pm$ 0.2 & 
1.485 $\pm$ 0.07 & 
0.356 $\pm$ 0.07 & 
0.0189 $\pm$ 0.002 & 
0.027 $\pm$ 0.0 & 
0.186 $\pm$ 0.02 & 
-0.228 $\pm$ 0.04 & 
0.415 $\pm$ 0.03 & 
-0.62 $\pm$ 0.16 & 
0.404 $\pm$ 0.08 & 
2.383 $\pm$ 0.17 \\
\cline{2-14} & Q High & 
-1.895 $\pm$ 0.19 & 
1.863 $\pm$ 0.19 & 
1.468 $\pm$ 0.06 & 
0.401 $\pm$ 0.09 & 
0.0185 $\pm$ 0.002 & 
0.025 $\pm$ 0.0 & 
0.194 $\pm$ 0.01 & 
-0.258 $\pm$ 0.13 & 
0.416 $\pm$ 0.04 & 
-0.654 $\pm$ 0.17 & 
0.429 $\pm$ 0.07 & 
2.293 $\pm$ 0.18 \\
\hline
\hline

\multirow{2}{*}{TUC} & Q Low & 
-1.887 $\pm$ 0.19 & 
1.846 $\pm$ 0.18 & 
1.457 $\pm$ 0.07 & 
0.407 $\pm$ 0.09 & 
0.0183 $\pm$ 0.002 & 
0.024 $\pm$ 0.0 & 
0.195 $\pm$ 0.01 & 
-0.26 $\pm$ 0.13 & 
0.415 $\pm$ 0.04 & 
-0.674 $\pm$ 0.19 & 
0.424 $\pm$ 0.07 & 
2.27 $\pm$ 0.18 \\
\cline{2-14} & Q High & 
-1.886 $\pm$ 0.2 & 
1.995 $\pm$ 0.19 & 
1.494 $\pm$ 0.06 & 
0.35 $\pm$ 0.07 & 
0.0191 $\pm$ 0.002 & 
0.028 $\pm$ 0.0 & 
0.185 $\pm$ 0.02 & 
-0.228 $\pm$ 0.06 & 
0.415 $\pm$ 0.04 & 
-0.602 $\pm$ 0.14 & 
0.407 $\pm$ 0.07 & 
2.402 $\pm$ 0.16 \\
\hline
\hline

\multirow{2}{*}{TFC} & Q Low & 
-1.831 $\pm$ 0.17 & 
1.922 $\pm$ 0.2 & 
1.461 $\pm$ 0.07 & 
0.381 $\pm$ 0.1 & 
0.019 $\pm$ 0.002 & 
0.026 $\pm$ 0.0 & 
0.189 $\pm$ 0.01 & 
-0.246 $\pm$ 0.13 & 
0.41 $\pm$ 0.03 & 
-0.633 $\pm$ 0.16 & 
0.397 $\pm$ 0.07 & 
2.319 $\pm$ 0.18 \\
\cline{2-14} & Q High & 
-1.937 $\pm$ 0.2 & 
1.923 $\pm$ 0.21 & 
1.489 $\pm$ 0.06 & 
0.375 $\pm$ 0.07 & 
0.0185 $\pm$ 0.002 & 
0.026 $\pm$ 0.0 & 
0.19 $\pm$ 0.02 & 
-0.241 $\pm$ 0.06 & 
0.42 $\pm$ 0.04 & 
-0.641 $\pm$ 0.17 & 
0.431 $\pm$ 0.07 & 
2.354 $\pm$ 0.18 \\
\hline
\hline

\multirow{4}{*}{TPC} & Q1 & 
-1.863 $\pm$ 0.19 & 
1.838 $\pm$ 0.18 & 
1.447 $\pm$ 0.07 & 
0.413 $\pm$ 0.11 & 
0.0186 $\pm$ 0.002 & 
0.024 $\pm$ 0.0 & 
0.195 $\pm$ 0.01 & 
-0.269 $\pm$ 0.17 & 
0.412 $\pm$ 0.03 & 
-0.661 $\pm$ 0.18 & 
0.417 $\pm$ 0.07 & 
2.255 $\pm$ 0.17 \\
\cline{2-14} & Q2 & 
-1.88 $\pm$ 0.2 & 
1.906 $\pm$ 0.2 & 
1.467 $\pm$ 0.07 & 
0.387 $\pm$ 0.07 & 
0.0187 $\pm$ 0.002 & 
0.025 $\pm$ 0.0 & 
0.192 $\pm$ 0.02 & 
-0.243 $\pm$ 0.07 & 
0.414 $\pm$ 0.03 & 
-0.623 $\pm$ 0.14 & 
0.416 $\pm$ 0.07 & 
2.322 $\pm$ 0.18 \\
\cline{2-14} & Q3 & 
-1.895 $\pm$ 0.2 & 
1.945 $\pm$ 0.2 & 
1.485 $\pm$ 0.06 & 
0.367 $\pm$ 0.07 & 
0.0187 $\pm$ 0.002 & 
0.027 $\pm$ 0.0 & 
0.188 $\pm$ 0.02 & 
-0.237 $\pm$ 0.06 & 
0.419 $\pm$ 0.04 & 
-0.642 $\pm$ 0.18 & 
0.416 $\pm$ 0.08 & 
2.36 $\pm$ 0.17 \\
\cline{2-14} & Q4 & 
-1.912 $\pm$ 0.19 & 
2.003 $\pm$ 0.19 & 
1.506 $\pm$ 0.05 & 
0.344 $\pm$ 0.07 & 
0.0189 $\pm$ 0.002 & 
0.028 $\pm$ 0.0 & 
0.184 $\pm$ 0.02 & 
-0.225 $\pm$ 0.04 & 
0.415 $\pm$ 0.03 & 
-0.617 $\pm$ 0.16 & 
0.414 $\pm$ 0.07 & 
2.417 $\pm$ 0.16 \\
\hline

\end{tabular}
\label{Scaling_descriptors_summaryStats_3}
\end{table}
\end{landscape}
\restoregeometry


\end{document}